
\input epsf

\global\newcount\meqno
\def\eqn#1#2{\xdef#1{(\secsym\the\meqno)}
\global\advance\meqno by1$$#2\eqno#1$$}
%
\global\newcount\refno
\def\ref#1{\xdef#1{[\the\refno]}
\global\advance\refno by1#1} \global\refno = 1
\voffset=0.1in
\vsize=7.in 
\hsize=5in 
\magnification=1200 
\tolerance 10000
%
%
%
\font\sevenrm = cmr7  

\def\bphi{\bar\Phi}
\def\bg{\bar g}

\def\gsim{ \,\, \vcenter{\hbox{$\buildrel{\displaystyle >}\over\sim$}}
 \,\,}

\def\hg{{\hat g}}

\medskip

\baselineskip=0.1cm
\medskip
\nobreak
\medskip

\baselineskip 12pt plus 1pt minus 1pt \vskip 2in
\medskip
\centerline{\bf SELF-CONSISTENT RENORMALIZATION GROUP FLOW} 
\vskip 24pt
\centerline{Sen-Ben Liao$^{1}$\footnote{$^\dagger$}{electronic address:
senben@phy.ccu.edu.tw} Chi-Yong Lin $^{2,3}$\footnote{$^\ddagger$}{electronic
address: lcyong@mail.ndhu.edu.tw} and Michael Strickland
$^4$\footnote{$^*$}{electronic address: mike@phys.washington.edu}} 
\vskip 12pt
\centerline{\it Department of Physics $^1$} \centerline{\it National
Chung-Cheng University} \centerline{\it Chia-Yi, Taiwan R. O. C.} 
\vskip 12pt
\centerline{\it Instituto De Fisica, Universidade De Sao Paulo $^2$}
\centerline{\it C.P.66.318, CEP 05315-970, Sao Paulo, Brazil} 
\vskip 12 pt
\centerline{\it Department of Physics $^3$} \centerline{\it National
Dong-Hwa University} \centerline{\it Hua-Lien, Taiwan R. O. C.} 
\vskip 12pt
\centerline{\it and} \vskip 12pt \centerline{\it Department of Physics $^4$}
\centerline{\it University of Washington} \centerline{\it Seattle, WA\ \
98195-1560\ \ \ U.S.A.} 
\vskip 1.2in
\vskip 24 pt \baselineskip 12pt plus 2pt minus 2pt \centerline{{\bf ABSTRACT}}
\medskip
\medskip

A self-consistent renormalization group flow equation for the scalar 
$\lambda\phi^4$ theory is analyzed and compared
with the local potential approximation. The
two prescriptions coincide in the sharp cutoff limit but differ with 
a smooth cutoff. The dependence of the 
critical exponent $\nu$ on the 
smoothness parameter and the field of expansion is explored. An optimization
scheme based on the minimum sensitivity principle is employed to ensure 
the most rapid convergence of $\nu$ with 
the level of polynomial truncation.

\vskip 24pt 
\vfill 
\noindent CCU-TH-00-02 \hfill October 2000 
\eject

\centerline{\bf I. INTRODUCTION}
\medskip
\nobreak \xdef\secsym{1.}\global\meqno = 1
\medskip
\bigskip

In recent years there has been a resurging interest in using the exact
renormalization group (ERG) formalism for investigating non-perturbative
phenomena such as QCD in extreme conditions, dynamical chiral symmetry breaking
and critical behavior of statistical systems. The ERG framework pioneered by
Wilson \ref\wilson, Wegner and Houghton \ref\wegner, and Polchinski
\ref\polchinski\ was based on the concept of momentum-space blocking
transformation under which the large momentum degrees of freedom are
coarse-grained. Decreasing the cutoff systematically then leads to a nonlinear
RG evolution equation which dictates how the effect of the irrelevant, short-distance
modes can be incorporated into the low-energy effective blocked action.

The ERG flow equation may be derived starting from the generating functional for the
connected Green's functions:
\eqn\gener{ e^{W_k[J]}=\int D[\phi]e^{-{1\over 2}\phi\cdot C_k^{-1}\cdot\phi -
S[\phi]+J\cdot\phi},}
where $C_k={{\tilde\rho_{k,\sigma}(p)}\over{p^2(1-\tilde\rho_{k,\sigma}(p))}}$
is an additive infrared (IR) cutoff, and $\tilde\rho_{k,\sigma}(p)$ is a
smearing function which approaches $\Theta(p-k)$ as the smoothness
parameter $\sigma\to \infty$. Upon varying the effective IR scale $k$
infinitesimally followed by a Legendre transformation,
$-W_k[J]+J\cdot\Phi={\widetilde S}_k[\Phi] +{1\over 2}\Phi\cdot C_k^{-1}\Phi$,
$\Phi={{\delta W_k}/{\delta J}}$, the RG equation for the blocked action
${\widetilde S}_k[\Phi]$ is obtained as:
\eqn\exac{\eqalign{ k{{\partial {\widetilde S}_k} \over {\partial k}} &=
-{1\over 2}{\rm Tr}\biggl[{1\over C_k}\Bigl(k{{\partial C_k} \over{\partial
k}}\Bigr)\cdot \Bigl(1+C_k\cdot{\delta^2{\widetilde
S}_k\over{\delta\Phi^2}}\Bigr)^{-1} \Biggr] \cr & =-{1\over 2}{\rm
Tr}\Biggl[{1\over{\tilde\rho_{k,\sigma} (1-{\tilde\rho_{k,\sigma}})}}
\Bigl(k{\partial{\tilde\rho_{k,\sigma}}\over{\partial k}} \Bigr)
\Bigl(1+{{\tilde\rho_{k,\sigma}}\over p^2 (1-{\tilde\rho}_{k,\sigma})}~
{\delta^2{\widetilde S}_k\over{\delta\Phi^2}}\Bigr)^{-1} \Biggr].}}
The above equation has been derived and extensively analyzed 
\ref\morris\ref\wetterich\ref\others.
Albeit being exact, it cannot be solved analytically and further 
approximation must be made for practical applications. 
Thus far the most reliable
prescription which retains the original non-perturbative features has been the
derivative expansion \ref\tmorris\ where the blocked action is written as: 
\eqn\deri{ {\widetilde S}_k[\Phi]=\int_x\biggl\{ {Z_k(\Phi)\over 2}
(\partial_{\mu}\Phi)^2+U_k(\Phi)+O(\partial^4)\biggr\},}
with $Z_k(\Phi)$ and $U_k(\Phi)$ being the wave-function renormalization
constant and the blocked potential, respectively \ref\sb. In the
$O(\partial^0)$ order, the so-called Local Potential Approximation (LPA)
\ref\nicoll\ where $Z_k(\Phi)$ is taken to be unity,
the RG evolution is characterized by
\eqn\rgu{ k{{\partial U_k(\Phi)} \over {\partial k}}= {1\over 2}\int_p
\Bigl(k{{\partial\tilde\rho_{k,\sigma}(p)}\over {\partial k}}
\Bigr)~{U''(\Phi)\over{ p^2+\tilde\rho_{k,\sigma}(p) U''_k(\Phi) }}.}
The equation can now be readily solved by numerical methods \ref\mike.

A further approximation to the RG flow is to expand $U_k(\Phi)$ in powers of $\Phi$
and then truncate the series at some order $M$ \ref\hasenfratz\ref\margaritis:
\eqn\uexpa{ U_k(\Phi)=\sum_{n=1}^M{
g_{2n}(k)\over{(2n)!}}~(\Phi-\Phi')^{2n},\qquad\qquad
g_{2n}(k)={{\partial^{2n} U_k}\over{\partial
\Phi^{2n}}}\Big\vert_{\Phi=\Phi'},}
thereby transforming the partial differential equation into a set of $M$ coupled
ordinary equations. Since the field expansion point $\Phi'$
is arbitrary, it may be chosen 
at the origin, the $k$-dependent minimum or any other non-vanishing field
values \ref\aoki. Thus, polynomial truncation not only
allows for a direct study of the flow of the coupling constants, numerical
implementation is also greatly facilitated. However, the drawback of such
truncated scheme is that it leads to spurious effects such as an
oscillation in the value of the critical exponents and the appearance of
complex eigenvalues at large $M$ \morris\margaritis. 
Therefore, it would be desirable to
employ a prescription which ensures an unambiguous identification of the 
relevant fixed points as well as the convergence of physical quantities.

In Ref.\ref\lps\ an optimization procedure based on the principle of 
minimum sensitivity
was proposed for measuring the critical exponent $\nu$. 
The parameter $\sigma$ was dialed until it reached a value
at which the minimum sensitivity condition ${d\nu\over dM}\vert_{\sigma}=0$ 
is satisfied. At the optimal $\sigma$, 
$\nu$ depends least sensitively on $M$ due to the
maximum suppression of the higher-order contributions, leading to
a maximum rate of convergence for $\nu$.
An optimized
scheme which retains only the relevant set of
operators and the leading few irrelevant operators to
characterize the RG flow can have non-trivial consequences in more
complicated systems such as fermionic or gauge theories. In particular, the
computation could be simplified enormously if certain classes of higher-order
Feynman diagrams are suppressed. 

In the present work we continue our search for an optimized RG prescription
at the level of LPA.
However, instead of following the conventional ERG approach as previously considered 
\lps, we start from
the one-loop contribution and introduce a ``multiplicative'' 
regulating smearing function $\tilde\rho_{k,\sigma}(p)$ 
in the momentum space integration:
\eqn\onere{ {\tilde U}^{(1)}_k(\Phi)={1\over
2}\int_p\tilde\rho_{k,\sigma}(p)~{\rm ln}~ \Bigl[1+{V''(\Phi)\over
p^2}\Bigr].}
The use of 
$\tilde\rho_{k,\sigma}(p)$ allows one to see clearly how each Feynman diagram is 
regularized. From the above equation, a self-consistent RG equation can be 
constructed by a differentiation with respect to $k$ followed by the 
substitution $V''(\Phi)\rightarrow U_k''(\Phi)$ on the right-hand-side:
\eqn\scrg{ k{{\partial U_k(\Phi)}\over{\partial k}}={1\over 2}\int_p
\biggl(k{\partial{\tilde\rho_{k,\sigma}(p)}\over{\partial k}}\biggr)
{\rm ln}\Bigl[1+{U_k''(\Phi)\over p^2}\Bigr].}
This procedure is analogous to the Schwinger-Dyson self-consistent improvement.
The main focus of this paper is to illustrate how Eq. \scrg\ can provide an
alternative, viable prescription that retains the 
non-perturbative information contained in the original full RG equation for
$\widetilde S_k[\Phi]$. In particular, the convergence of 
physical quantities such as the critical exponents extracted using polynomial
truncation will be demonstrated.

The organization of the paper is
as follows. In Sec. II we briefly review the essential features of LPA and
compare with the self-consistent RG equation, Eq. \scrg.
The sharp cutoff limit of the RG flow 
is discussed in Sec. III and the dependence of the critical exponent $\nu$ on the 
field expansion point $\Phi_0$ and the level of truncation $M$ is also explored. 
We expand the potential about its $k$-dependent
minimum, a prescription which does not respect the $Z_2$ symmetry of the original  
flow equation. Nevertheless, the resulting $\nu$ converges remarkably with increasing
$M$ up to $M=26$.  
In Sec. IV the RG equations generated by
smooth smearing functions are presented. Unlike the sharp cutoff case, 
singularities which resemble spontaneous symmetry breaking
are present in the self-consistent integro-differential equation for the 
fixed-point solution ${\bar U}^{*}(\bphi)$ when employing a smooth cutoff. 
To overcome the complication we again make use of the 
non-invariant expansion and obtain
numerical solutions for $\nu$ up to $M=20$.
Sec. V is reserved for summary and discussions. In Appendix A we
summarize some detail associated with polynomial truncation
of the sharp flow equation. The manner in which propagators are modified by LPA and 
the self-consistent RG schemes is discussed in Appendix B. In
Appendix C we explore the large $\sigma$ limit of the self-consistent RG. 

\bigskip
\medskip

\centerline{\bf II. LPA AND SELF-CONSISTENT RG}
\medskip
\nobreak \xdef\secsym{2.}\global\meqno = 1
\medskip
\nobreak

We first give a brief recapitulation of the basic features of LPA. The
ERG equation can be constructed by
introducing a smearing function $\rho_{k,\sigma}(x)$ which governs the
coarse-graining procedure and $k$ acts as an effective IR cutoff, separating
the low- and the high-momentum modes. The propagator is modified as:
\eqn\smprop{ \Delta(p)={1\over p^2}\longrightarrow \Delta_{k,\sigma}(p)
={{1-\rho_{k,\sigma}(p)}\over p^2}={{\tilde\rho_{k,\sigma}(p)}\over p^2},}
where $\rho_{k,\sigma}(p)+\tilde\rho_{k,\sigma}(p)=1$. In LPA where
$Z_k(\Phi)=1$ the evolution is characterized by the flow of $U_k(\Phi)$. 
For a generic smooth function $\rho_{k,\sigma}(p)=
\Theta_{\sigma}(k,p)$, it is
convenient to write $\Theta_{\sigma}(k,p)=t$ where $0 \le t \le 1$, and 
the reparameterization leads to
\eqn\rgt{ k{{ dU_k(\Phi)} \over dk} = {S_d\over
2}\int_{t(p=0)}^{t(p=\infty)}dt~ {p^d(t)~U''(\Phi)\over{
p^2(t)+(1-t)U''_k(\Phi)}}.}
Thus, by inverting $\rho_{k,\sigma}(p)$
for $p(t)$, the RG equation of $U_k(\Phi)$ is readily obtained.

Clearly, the evolution equation depends on the
choice of the smearing function. While a sharp cutoff 
provides a well-defined boundary between the high (integrated) and the low
(unintegrated) momentum modes and yields a partial differential
equation, no clear separation exists for a smooth cutoff and the
RG equation retains its integral-differential form. 
In principle, physical
content extracted from solving the RG flow should not depend on the shape of the
cutoff which merely reflects how the irrelevant degrees of
freedom are integrated over. However, due to the approximation employed
such dependence is explicitly present and the critical exponent $\nu$
exhibits a small dependence on $\sigma$ \tmorris. The dependence may
indicate that the effect of certain relevant operators has been 
neglected \ref\litim.

Within the framework of LPA, we see that Eq. \rgu\ may
also be obtained by a similar self-consistent ``trick'' starting from 
\ref\lp:
\eqn\oneloop{ {\tilde U}^{(1)}_k(\Phi)={1\over 2}\int_p{\rm ln}~
\Bigl[1+\tilde\rho_{k,\sigma}(p){V''(\Phi)\over p^2}\Bigr].}
This can be understood by noting
that since we are now using an ``exact'' propagator, 
all the higher-order Feynman graphs 
that belong to the one-loop resummed class (e.g., daisy and super-daisy)
are automatically incorporated. More generally, we may write:
\eqn\onel{ {\tilde U}^{(1)}_k(\Phi)={1\over 2}\int_p{\rm ln}~ \Bigl[1+{\cal
P}_{k,\sigma}(p){V''(\Phi)\over p^2}\Bigr],}
with
\eqn\efpr{ {\cal P}_{k,\sigma}(p)~{1\over p^2}={\tilde\rho_{k,\sigma}(p)\over
p^2} =\Delta_{k,\sigma}(p).}
Here ${\cal P}_{k,\sigma}(p)$ acts as a projection operator on the
massless propagator $1/p^{2}$ and turns it into an effective propagator
$\tilde\rho_{k,\sigma}(p)/p^2$ which contains only the high momentum modes $p
\gsim k$. For example, in the sharp limit one would have
$\Delta_k(p)=\Theta(p-k)/p^2$.

Let us examine how the alternative, self-consistent RG prescription based on 
Eq. \onere\  differ from the usual LPA. First we
note that since the coarse-graining procedure is initiated in the high-momentum
regime where $V''(\Phi)/p^2 \ll 1$, Eq. \oneloop\ may be expanded as
\eqn\onelo{\eqalign{ {\tilde U}^{(1)}_k(\Phi) &={1\over 2}\int_p
\sum_{n=1}^{\infty}{(-1)^{n+1}\over n} \biggl[{\cal P}_{k,\sigma}(p)~{V''(\Phi)\over
p^2}\biggr]^n ={1\over 2}\int_p \sum_{n=1}^{\infty}{(-1)^{n+1}\over
n}\biggl[\tilde\rho_{k,\sigma} {V''(\Phi)\over p^2}\biggr]^n \cr 
& 
={1\over2}\int_p~{\tilde\rho_{k,\sigma}}~{\rm ln}\Bigl[1+{V''(\Phi)\over p^2}\Bigr] 
+{1\over 2}\sum_{n=1}^{\infty}{(-1)^{n+1}\over n+1}~\int_p~{\tilde\rho_{k,\sigma}}\bigl(
1-\tilde\rho_{k,\sigma}^n\bigr)\Bigl({V''(\Phi)\over p^2}\Bigr)^{n+1}.}}
Differentiating both sides with respect to ${\rm ln}\,k$ and invoking the self-consistent
prescription then leads to
\eqn\rgty{\eqalign{ k{{\partial U_k(\Phi)} \over {\partial k}} &={1\over
2}\int_p \Bigl(k{{\partial \tilde\rho_{k,\sigma}(p)}\over {\partial k}}\Bigr)~
{\rm ln}\Bigl[1+{U_k''(\Phi)\over p^2}\Bigr] 
\cr 
& +{1\over
2}\sum_{n=1}^{\infty}{(-1)^{n+1}\over n+1}\int_p \Bigl(k{{\partial
\tilde\rho_{k,\sigma}(p)}\over {\partial k}}\Bigr)~
\Bigl[1-(n+1)\tilde\rho_{k,\sigma}^n(p)\Bigr]\Bigl({U_k''(\Phi)\over
p^2}\Bigr)^{n+1}.}}
In the first term of the equation above one finds a complete factorization
between the running of the scale $k$ and the $\Phi$ dependence. This is the 
equation which we shall focus on in the subsequent sections.
The second term, representing the difference between the self-consistent RG
and the LPA, can also be written as:
\eqn\rgtd{ \Delta U_k(\Phi) ={S_d\over
2}\int_{t(p=0)}^{t(p=\infty)}dt~p^d(t)\Biggl\{ {U_k''(\Phi)
\over{p^2(t)+(1-t)U_k''(\Phi)}}- {\rm ln}\Bigl[ p^2(t)+U''_k(\Phi)\Bigr]\Biggr\}.}
From Eq. \onelo\ or \rgtd, we see that the two RG schemes coincide
when $\tilde\rho_{k,\sigma}(p)\to 1$ or 0. That is,
$\Delta U_k(\Phi)=0$ when ${\cal P}_{k,\sigma}^2(p)={\cal P}_{k,\sigma}(p)$ or
$\tilde\rho_{k,\sigma}^2(p)=\tilde\rho_{k,\sigma}(p)$. However, this holds only
for the trivial case of a sharp cutoff; smooth cutoffs such as
$\tilde\rho_{k,b}(p)=1-e^{-a(p/k)^b}$ and
$\tilde\rho_{k,m}(p)=(p/k)^m/(1+(p/k)^m)$, on the other hand, give a small
deviation for large $m$ and $b$ in the regime $p/k \gg 1$.

The salient feature of our approach is that once the perturbative results are known, 
they can be readily transformed into the corresponding self-consistent RG equations. 
Generalization
to higher order derivative expansion can be achieved with ease. 
For example, at order $O(\partial^2)$, we have:
\eqn\uex{ k{{\partial U_k(\Phi)}\over{\partial k}}={1\over 2}\int_p
\biggl(k{\partial{\tilde\rho_{k,\sigma}(p)}\over{\partial k}}\biggr)
~{\rm ln}\Bigl[1+{U_k''(\Phi)\over Z_k(\Phi)p^2}\Bigr],}
and \ref\alexandre\ref\enzo\
\eqn\zex{\eqalign{
k{{\partial Z_k}\over{\partial k}} 
&= \int_p \biggl(k{\partial{\tilde\rho_{k,\sigma}(p)}\over{\partial k}}
\biggr)\Biggl\{ {Z_k''\over{Z_kp^2+U_k''}}
-2Z'_k~{{Z'_kp^2+U_k'''}\over (Z_kp^2+U_k'')^2} 
-{p^2\over d}~{(Z'_k)^2\over (Z_kp^2+U''_k)^2} \cr
&
+{4p^2\over d}~Z_kZ'_k{{Z_k'p^2+U'''_k}\over (Z_kp^2+U''_k)^3}
+Z_k{{(Z_k'p^2+U'''_k)^2}\over (Z_kp^2+U''_k)^3}
-{4p^2\over d}~Z_k^2{(Z'_kp^2+U_k''')^2\over (Z_kp^2+U''_k)^4}\Biggr\}.}}
We comment that the above prescription, being reminiscent to the Schwinger-Dyson 
approach, has been widely employed to treat more complicated issues such as 
the effective quark propagator \ref\mthesis\ and dynamical chiral symmetry 
breaking in gauge theories \ref\chiral. 
One great advantage here is that, unlike the exact RG approach, 
the flow for $Z_k(\Phi)$ is completely analytic, even in the 
sharp cutoff limit, and the numerical algorithm can be readily implemented.
Incorporating the effect of $Z_k(\Phi)$ with the above approach 
has been shown to improve the accuracy of $\nu$ \ref\wambach.
In what follows we shall set $Z_k(\Phi)=1$ ($\eta=0$) for simplicity.

Although we shall mainly be concerned with smearing functions 
that approach a step function as $\sigma\to\infty$ in our self-consistent RG scheme, 
other type of regularization can also be utilized.
For example, one could choose the smearing function to be of the Pauli-Villars type: 
\eqn\pvi{ {\tilde\rho}_k(p)=\Bigl({\Lambda^2\over{p^2+\Lambda^2}}\Bigr)^{d/2}
-\Bigl({k^2\over{p^2+k^2}}\Bigr)^{d/2}.}
In a similar manner, the spirit of the dimensional regularization can 
also be encapsulated
by using ${\tilde\rho}_{\epsilon}(p)={d^{-\epsilon}p}/(2\pi)^{-\epsilon}$ in integer 
$d$; with $\lambda\to \lambda\mu^{\epsilon}$, the RG flow is described 
by $\mu$, the renormalization point.

\bigskip
\medskip

\centerline{\bf III. SELF-CONSISTENT RG - SHARP CUTOFF}
\medskip
\nobreak \xdef\secsym{3.}\global\meqno = 1
\medskip
\nobreak

Before considering smooth cutoffs, it is instructive to first
review the sharp cutoff results \lps. 
In order to recover this limit, caution must be taken to
invert $t=\Theta(1-z)$. Here it must be interpreted that $t$ changes {\it
continuously} from $1$ to $0$ at $z=1$, thereby leading to \morris:
\eqn\drgti{ \Biggl[k{\partial\over{\partial k}}-{1\over 2}\bigl(d-2)
\bar\Phi{\partial\over{\partial\bar\Phi}}+d\Biggr]{\bar U}_k(\bar\Phi)
=-\int^1_0dt~{\rm ln}\Bigl[1+{\bar U}''_k(\Phi)\Bigr] =-{\rm ln}~\Bigl[1+{\bar
U}_k''(\Phi)\Bigr],}
where ${\bar U}_k(\bar\Phi)=\zeta^2k^{-d}U_k(\Phi)$, $\bar\Phi=\zeta
k^{-(d-2)/2} \Phi$, with $\zeta=\sqrt{2/S_d} =\sqrt{(4\pi)^{d/2}\Gamma(d/2)}$.
The fixed-point solution is given by
\eqn\fpflow{ -{1\over 2}\bigl(d-2)\bar\Phi {\bar U}^{*'}+ d{\bar U}^{*}
=-{\rm ln}~[1+{\bar U}^{*''}],}
with $U^{*'}(0)=0$ from reflection symmetry. 

However, it has been pointed out that  
only certain values of $\bg_0^{*}={\bar U}^{*}(0)=-{\rm ln}(1+\bg_2^{*})/d$ 
can lead to physically meaningful solutions \morris\hasenfratz; in all other 
cases a singularity 
of the form ${\bar U}^{*}\sim {\rm ln}({\bar\Phi}_c-{\bar\Phi})$ will be  
encountered and the potential becomes 
complex beyond some critical value $\bar\Phi_c$.
In $d=3$, only two solutions have been found; the first one is $\bg_0^{*}=0$ which
corresponds to the trivial Gaussian
fixed point and the second one, $\bg_0^{*}=0.206343\equiv s_0$~
(or $\bg_2^{*}=-0.461533\equiv s$), 
the Wilson-Fisher. The qualitative feature of the fixed-point solutions
is illustrated in Fig. 1.

\medskip
\medskip

\centerline{\epsfbox{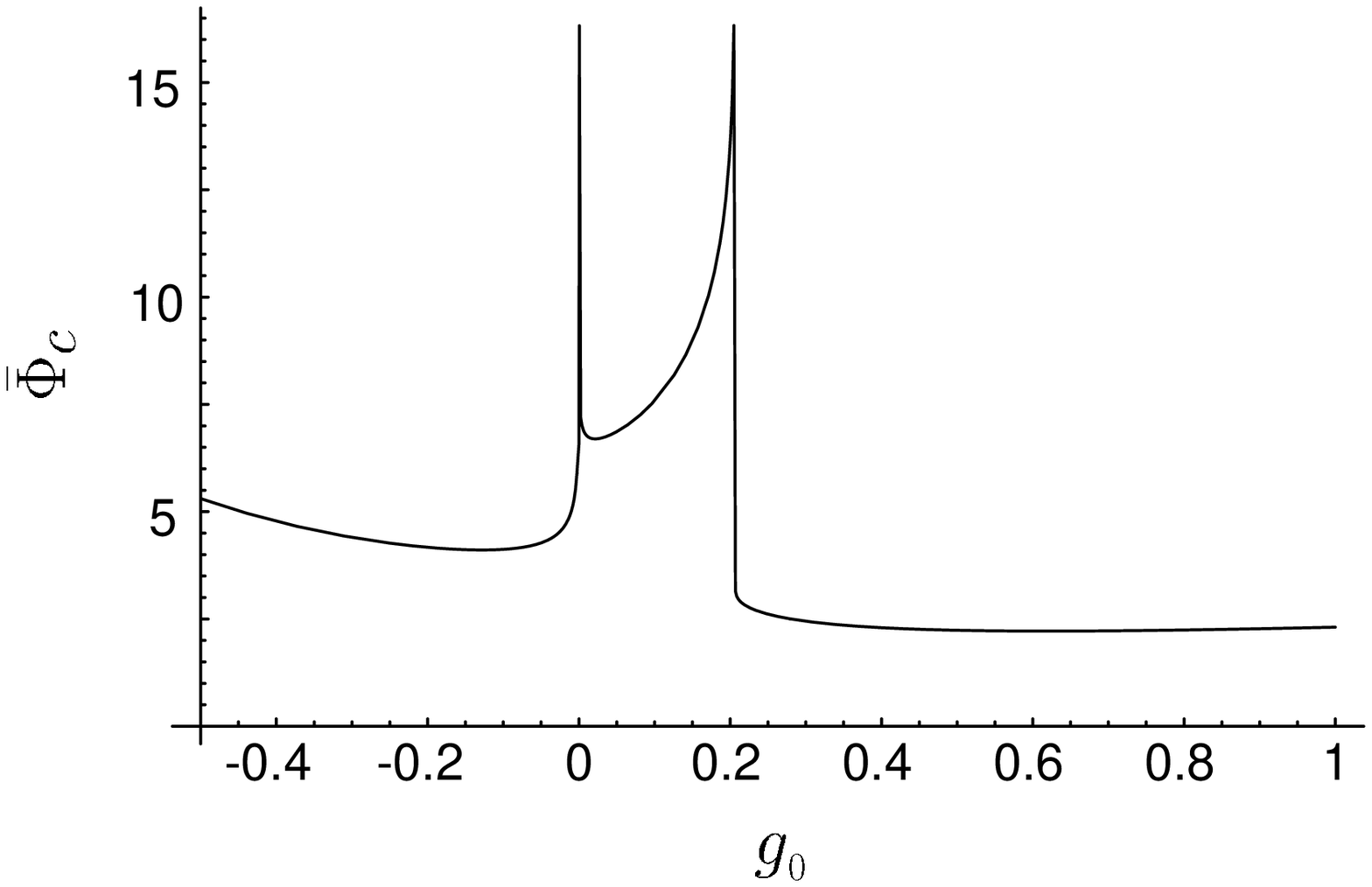}}
\medskip
{\narrower
{\sevenrm
{\baselineskip=8pt
\itemitem{Figure 1.}
Singularity $\scriptstyle {\bar\Phi}_c$ as a function of $\scriptstyle g_0$
in $\scriptstyle d=3$. Notice that there are only two physical solutions, one
for the Gaussian and the other one for the Wilson-Fisher fixed point.
\bigskip
}}}


As we mentioned in the Introduction, instead of solving the full flow equation,
we may expand ${\bar U}_k(\bar\Phi)$ in powers of $\bar\Phi$ and terminate the
series at $\bar\Phi^{2M}$, thereby turning the nonlinear partial differential
equation into a set of
$M$ coupled ordinary differential equations. 
We consider below expansion schemes that are both $Z_2$-invariant and non-invariant.
\medskip
\medskip
\bigskip

\centerline{\bf a. invariant expansion}
\medskip

The simplest invariant polynomial expansion of 
${\bar U}_k(\bar\Phi)$ can be carried out around $\bar\Phi=0$, as in Eq. \uexpa. 
The $\bar\beta$ functions, $\bar\beta_n=k{{\partial {\bar
g}_n}/{\partial k}}$, can be written as:
\eqn\snwo{\eqalign{\bar\beta_2 &=-2{\bar g}_2-{\bar G}_4, \cr 
\bar\beta_4 &=-\epsilon{\bar g}_4-\bigl(-3{\bar G}_4^2+{\bar G}_6\bigr), \cr 
\bar\beta_6 &=(2d-6){\bar g}_6-\bigl(30{\bar G}_4^3-15{\bar G}_4{\bar G}_6
+{\bar G}_8\bigr),\cdots}}
where ${\bar G}_n={\bar g}_n/(1+{\bar g}_2)$ \ref\jean, and the fixed points can
now be located by setting $\bar\beta_n=0$ for all $n$. 
If only ${\bar
g}_2$ and ${\bar g}_4$ are kept, we find, besides the trivial fixed point $({\bar
g}_2^{*}, {\bar g}_4^{*})=(0,0)$, an additional Wilson-Fisher fixed point
$({\bar g}_2^{*},{\bar g_4}^{*})= \bigl(-{\epsilon\over {6+\epsilon}},
~{12\epsilon\over{(6+\epsilon)^2}}\bigr)$.
Linearizing the flow about the above fixed point, the
eigenvalues can easily be found to be
$[(-3+2\epsilon)\pm\sqrt{9+6\epsilon+7\epsilon^2}]/3$, or $(-1\pm\sqrt{22})/3$ for
$\epsilon=1$. By our convention, the 
largest {\it negative} eigenvalue corresponds to the inverse of the critical
exponent $1/{\nu}$. Thus, we have $\nu \approx 0.5272$ at $M=2$. We remark that in the
spirit of the conventional $\epsilon$-expansion where $\epsilon$ is first taken to be
``small,'' the matrix element ${\partial{\bar\beta}_4}/{\partial{\bar g}_2}$
which is $O(\epsilon^2)$ is discarded, and 
$\nu=0.6$ instead \ref\zinn. Thus, when characterizing the flow by powers
of $\bar\Phi$, there is a reshuffling of contributions from the
$\epsilon$ series. 

\medskip
\medskip

\centerline{\epsfbox{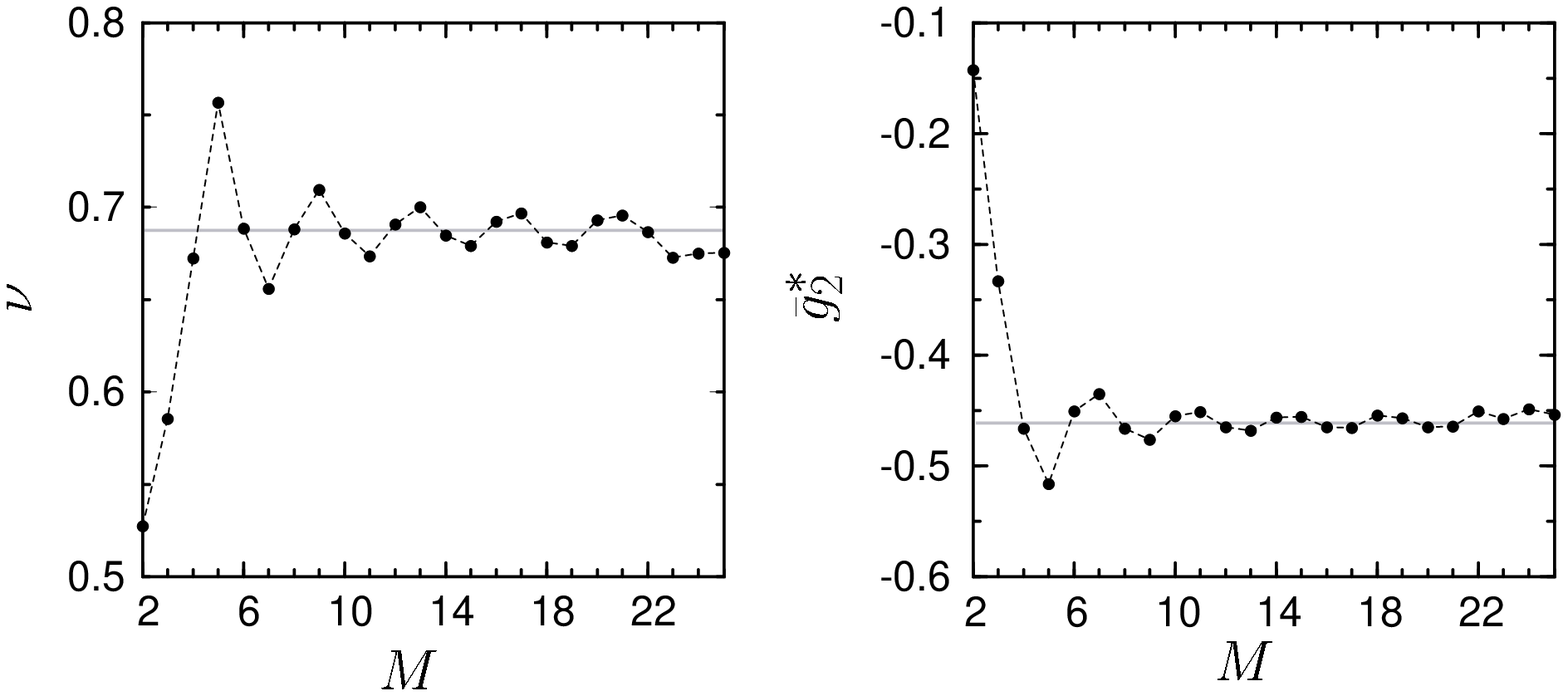}}
\medskip
{\narrower
{\sevenrm
{\baselineskip=8pt
\centerline{{Figure 2.}
Oscillation of critical exponent $\scriptstyle \nu$  and $\scriptstyle g_2^{*}$ 
as a function of $\scriptstyle M$.}
\bigskip
}}}

\medskip
\medskip

However, it is well-known that $\nu$  obtained via polynomial 
truncation with a sharp cutoff generally does not converge but oscillates with 
$M$, as depicted in Fig. 2. We find that at $M=22$ an
additional relevant direction is generated; and as $M$ is further increased,
more spurious solutions are found and the
identification of the Wilson-Fisher fixed point becomes ambiguous. 
The oscillation in $\nu$ is intimately linked to the oscillation in 
$\bg_2^{*}$ and the cause is related to the singular structure of 
the non-truncated fixed point potential  
${\bar U}^{*}(\bar\Phi)$ on the complex plane \morris; namely, the asymptotic
behavior is dominated by a pole 
$\bar\Phi_c=|{\bar\Phi}_c|e^{i\theta_c}$ having $\theta_c\approx
\pm \pi/2$ that gives rise to an approximate 
four-fold periodicity.  
Such truncation dependence is not unexpected since 
the fixed point(s) one identifies at each order is only approximate
and not the {\it true} Wilson-Fisher fixed point; therefore,
the would-be irrelevant set of operators will continue to evolve, leading to 
the $M$ dependence of physical quantities. 
In order to improve the convergence of $\nu$ in
polynomial truncation, Aoki {\it et al} has carried out an expansion
about a $Z_2$-invariant field $\bar\chi_0=\bphi_0^2/2$ \aoki:
\eqn\ueo{ {\bar U}_k(\bar\chi)=\sum_{n=0}^{M}{
c_{n}(k)\over{n !}}~(\bar\chi-\bar\chi_0)^{n},\qquad\qquad
c_{n}(k)={\bar U}^{(n)}_k(\chi_0)={{\partial^{n} {\bar U}_k}\over{\partial
\bar\chi^{n}}}\Big\vert_{\bar\chi=\bar\chi_0}.}
The convergence has been found to improve dramatically.
A more thorough discussion of this expansion scheme is presented in Appendix A. 
\medskip
\medskip
\centerline{\bf b. non-invariant expansion}
\medskip

Alternatively, one may expand the 
potential around some non-vanishing $\bphi'$:
\eqn\uem{ {\bar U}_k(\bar\Phi)=\sum_{n=0}^{M}{
{\hat g}_{n}(k)\over{n !}}~(\bar\Phi-\bar\Phi')^{n},\qquad\qquad
{\hat g}_{n}(k)={\bar U}^{(n)}_k(\Phi')={{\partial^{n} {\bar U}_k}\over{\partial
\bar\Phi^{n}}}\Big\vert_{\bar\Phi=\bar\Phi'}.}
Here the presence of odd powers of $\bphi$ will explicitly break the $Z_2$
symmetry. However, this novel
prescription becomes particularly useful when spontaneous symmetry breaking 
takes place,
or when the running mass parameter ${\bar g}_2(k)$ is still negative during the 
RG evolution. It also allows us to monitor the flow around $\bar\Phi'$
instead of the origin. 

Choosing $\bar\Phi'=\bar\Phi_0$, the $k$-dependent minimum and using Eq. \uem, 
we obtain
\eqn\nwt{\eqalign{ \dot{\bar\Phi}_0 &= -(1-{\epsilon\over 2}){\bar\Phi}_0
+{\hat G}_3/{\hg_2},\cr
\hat\beta_2 &=\hg_3\dot{\bphi}_0-2{\hat g}_2+(1-{\epsilon\over 2})\bar\Phi_0\hg_3
+\bigl({\hat G}_3^2-{\hat G}_4\bigr), \cr
\hat\beta_3 &=\hg_4\dot{\bphi}_0-\bigl(1+{\epsilon\over 2}\bigr){\hat g}_3
+(1-{\epsilon\over 2})\bar\Phi_0\hg_4+
\bigl( -2{\hat G}_3^3+3{\hat G}_3{\hat G}_4-{\hat G}_5\bigr),\cr
\hat\beta_4 &=\hg_5\dot{\bphi}_0-\epsilon {\hat g}_4+(1-{\epsilon\over 2})
\bar\Phi_0\hg_5+\bigl( 6{\hat G}_3^4-12{\hat G}_3^2{\hat G}_4+3{\hat G}_4^2
+4{\hat G}_3{\hat G}_5 -{\hat G}_6\bigr),\cdots}}
where ${\hat G}_n={{\hat g}_n\over{1+{\hat g}_2}}$. 
At the fixed point where $\dot{\bphi}_0$
vanishes, we have $\bar\Phi_0^{*}={2{\hat g}_3^{*}\over{(2-\epsilon)
{\hat g}_2^{*}(1+{\hat g}_2^{*})}}$.
When solving the coupled equations ${\hat\beta}_n=0$ to locate the 
fixed points, one finds that the number of spurious solutions 
again increases with $M$. Moreover, due to the loss of $Z_2$ symmetry, 
the Wilson-Fisher fixed point no longer has just one relevant eigen-direction, 
as would be in the $Z_2$-symmetric case.

The new Wilson-Fisher fixed point which dictates the critical behavior
of the Ising universality class satisfies the following conditions:
(i) all the fixed-point coupling constants $\hg_n^{*}$'s are real, (ii) 
the extremum of ${\bar U}^{*}(\bphi)$ is close to $\bphi_0^{*}=1.9287$, 
the field value obtained from solving the full non-truncated LPA equation, 
and (iii) at $M=2j ~(j > 2)$  
there are $\ell=j+1$ relevant eigen-directions which,  
by our convention, have negative real part for the eigenvalues.
Let us see how the above criteria work. 
First we note that $\bg_2^{*} < 0$ for $M < 6$, an indication that
$\bphi_0$ is a local maximum. At $M=4$ there are only three non-trivial solutions
$(\bg_2^{*},\bg_3^{*},\bg_4^{*})=(-{1\over 7},0,{12\over 49})$ and
$(-0.4125,\pm 0.2327,0.4456)$, with the first being the usual $Z_2$-symmetric 
solution at $\bphi_0=0$. We choose the asymmetric solution with $\bg_3^{*} < 0$,
which yields 
$\bphi_0=1.9203$ and $\nu=0.4229$. For $M \ge 6$, the extremum becomes a
local minimum and we have $\ell=j+1$. This is due to the fact that
the fields that break the $Z_2$ symmetry remain relevant throughout. 
In addition, there are two more coupling constants, 
$\bg_2^{*}$ and $\bg_4^{*}$, which are relevant by power counting.
For example, at $M=10$ the relevant set should be
$\{\bg_2^{*},\bg_3^{*},\bg_4^{*},\bg_5^{*},\bg_7^{*}, \bg_9^{*}\}$, i.e., 
$\ell=6$. Note that the classification of operators around the trivial 
Gaussian fixed point is still based on the counting of their canonical dimensions
$[\bg_n]=d-n({d\over 2}-1)$. 

The fixed-point potential ${\bar U}^{*}(\bar\Phi)$ is plotted in Fig. 3 as
a function of $\bar \Phi$ for various $M$. Due to the  
$Z_2$ asymmetry and the choice $\bphi_0 > 0$ as the expansion point,
only $\bar U^{*}(\bar\Phi)$ with positive $\bphi$ is tracked.
We note that the approximation first improves with increasing $M$ 
and nearly coincides with
the full non-truncated potential at $M=20$. Incidentally
this is the order at which
the term linear in $\bar\Phi$ that represents the leading symmetry-breaking
contribution is minimized.
However, beyond $M=20$ the potential begins to deviate away from the 
non-truncated LPA solution.
At $M=28$ a sensible result can be obtained only by choosing a complex eigenvalue. 
The remarkable convergence of $\nu$ with increasing $M$ up to $M=26$ is 
depicted in Fig. 4.  
Note that in the $Z_2$-invariant polynomial truncation scheme, 
depending on the point of expansion, 
$\nu$ is found to converge to or oscillate about $0.689$, which is the value obtained 
by solving the non-truncated LPA equation. On the other hand, our
non-invariant prescription gives $0.649$ for a large range of $M$.
Such discrepancy
is not unexpected since the symmetry of fixed points 
for the Ising universality class differs in the two cases.
It is interesting also to note that the result is very close to the value $0.651$
obtained via the optimization procedure in which the most optimal smoothness
parameter is sought to achieve minimum sensitivity \lps.

\medskip
\medskip

\centerline{\epsfbox{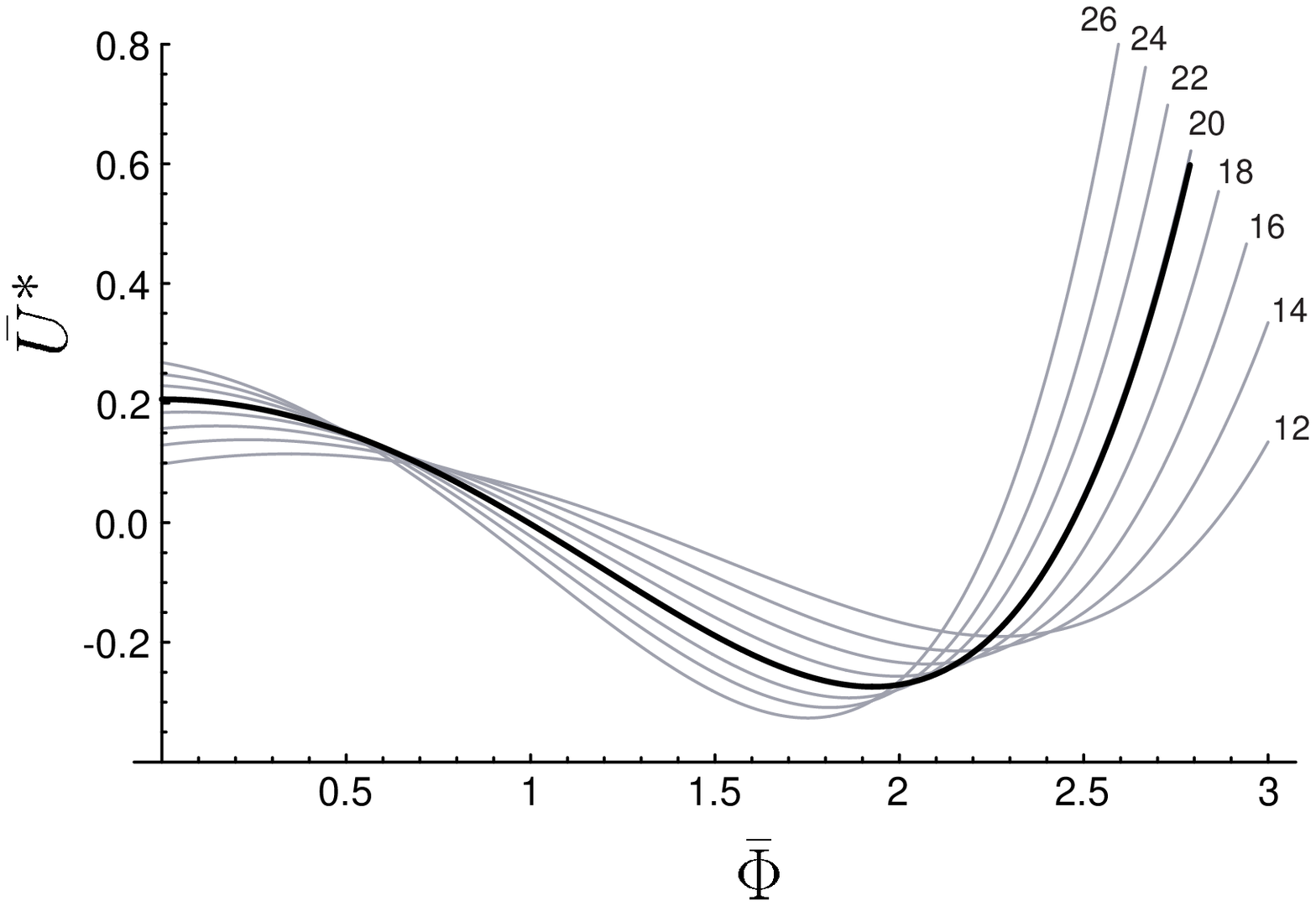}}
\medskip
{\narrower
{\sevenrm
{\baselineskip=8pt
\itemitem{Figure 3.} Fixed-point potential $\scriptstyle {\bar U}^{*}(\bar\Phi)$
obtained by polynomial truncation.
Exact LPA solution is shown as solid black line.
The agreement with the full LPA solution first
improves with increasing ${\scriptstyle M}$ and nearly coincides at 
$\scriptstyle M=20$.
\bigskip
}}}


\medskip
\medskip

\centerline{\epsfbox{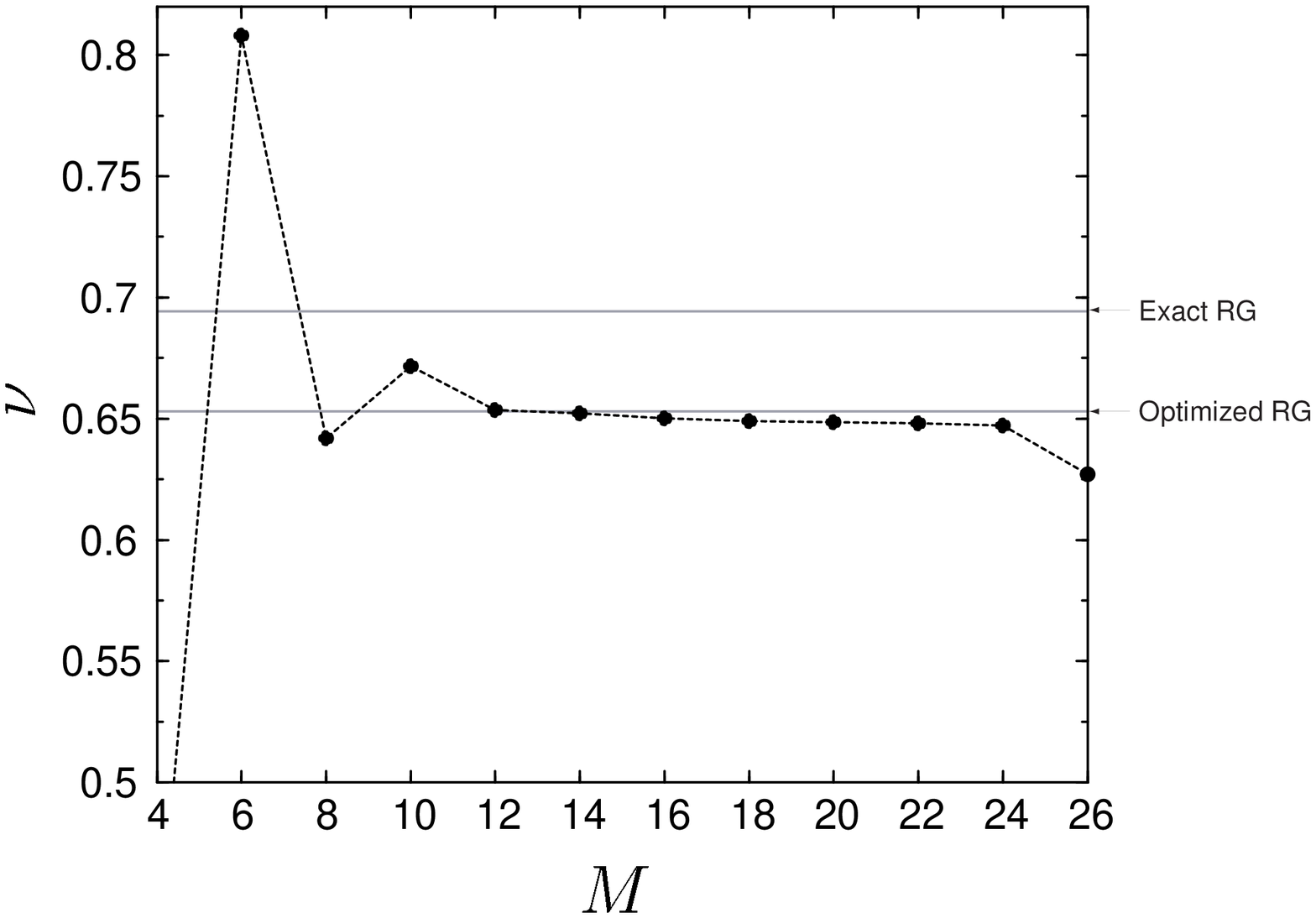}}
\medskip
{\narrower
{\sevenrm
{\baselineskip=8pt
\itemitem{Figure 4.}
Critical exponent $\scriptstyle \nu$ as a function of $\scriptstyle M$ obtained by
an non-invariant expansion of $\scriptstyle {\bar U}^{*}(\bar\Phi)$ about the minimum 
$\scriptstyle {\bar\Phi}_0$.
\bigskip
}}}


We comment that besides the origin or the $k$-dependent minimum $\bar\Phi_0$,
in principle $\bar\Phi'$ can be arbitrarily chosen. 
However, the rate of convergence is closed linked to the choice of $\bar\Phi'$.
Aoki {\it et al} has shown that expanding the sharp cutoff RG equation for
${\bar U}^{*}(\bar\Phi)$ about the minimum, $\bar\Phi_0$, yields a larger
radius of convergence compared to at the origin \aoki. In fact, it leads to
the exact solution in the large $N$ limit. Thus, when expanding about the
minimum $\bar\Phi_0$, a sharp cutoff smearing function with $\sigma\to\infty$
would be the optimal ``smoothness'' that satisfies the minimum sensitivity
principle, as we shall see in the next Section. 

\bigskip
\medskip

\centerline{\bf IV. SELF-CONSISTENT RG  - SMOOTH CUTOFFS}
\medskip
\nobreak \xdef\secsym{4.}\global\meqno = 1
\medskip
\nobreak

In this Section we examine the fixed-point potential ${\bar U}^{*}(\bar\Phi)$ 
obtained via polynomial truncation with smooth cutoffs.
Unlike the sharp cutoff case, the RG flow is characterized by the following 
(dimensionless) integro-differential equation:
\eqn\nrgta{\eqalign{\Biggl[k{\partial\over{\partial k}}-{1\over 2}\bigl(d-2)
\bar\Phi{\partial\over{\partial\bar\Phi}}+d\Biggr]{\bar U}_k(\bar\Phi)
&=\int_0^{\infty}dz~z^{d-1}\Bigl(k{{\partial \tilde\rho_{k,\sigma}}\over
{\partial k}}\Bigr)~{\rm ln}\Bigl[z^2+ {\bar U}_k''(\bar\Phi)\Bigr] \cr 
&
=-\int_0^1dt~z^d(t)~{\rm ln}\Bigl[z^2(t)+{\bar U}_k''(\bar\Phi)\Bigr].}}
The smearing functions are chosen in such a way that the conditions
$t(z=0)=1$ and $t(z=\infty)=0$ are always satisfied. 
However, we shall find that within a certain range of $k$ in the smooth case
the running mass parameter $\bg_2(k)$ (as well as $\bg_2^{*}$)
are negative quantities and singularity is encountered in $\bar\beta_n$ 
as the denominator $z^2(t)+{\bar U}''_k(\bphi)$ goes to zero. 

The reason why such 
singularity is absent in the sharp cutoff but present in the smooth cutoff 
for self-consistent
RG can be understood from the manner in which the large 
momentum modes are eliminated.
In the case of exact RG a sharp cutoff provides a clear 
separation between the integrated $(z\equiv p/k > 1)$ and the unintegrated 
modes $(z\equiv p/k < 1)$. With a negative mass curvature 
$\bg_2^{*}= s \approx -0.461533$ (in unit of $k$) for the non-truncated potential,
the integration 
is completely analytic because the condition $1+s > 0$ is always satisfied in the 
denominator.
However, in the self-consistent RG formulation no sharp boundary 
exists for a smooth cutoff and for any finite 
$\sigma$ the range of the $z$ integration is stretched from
zero to infinity. Therefore, singularity sets in
when $z^2+s \le 0$. The situation confronted here actually resembles
that of spontaneous symmetry breaking where the potential is also double-welled. 
However,
we emphasize that ${\bar U}^{*}(\bphi)$ describes a scale-invariant theory in the 
symmetric phase; as we integrate down to $k=0$ in the {\it dimensionful} RG equation, 
the potential will become single-welled with a unique minimum at the origin \mike. 
True symmetry breaking is marked by $\lim_{k\to 0}g_2(k) < 0$. 
Notice that the
integration with a smooth cutoff would be completely analytic had the LPA prescription
been employed; the difference between the two schemes may be attributed to the 
manner in which the propagators are modified. 
The details are further discussed in Appendix B.

To overcome the difficulty associated with the divergence near $z=\sqrt{-\bg_2^{*}}$, 
we expand the potential about a non-zero minimum $\bphi_0$ where 
${\bar U}''(\bphi_0)=\hg_2 > 0$, as in Eq. \uem. In analogy to Eq. \nwt, we have,
for the smooth case:
\eqn\nmin{\eqalign{\dot{\bphi}_0 &=-\bigl(1-{\epsilon\over 2}\bigr)\bphi_0+\int_0^1~
z^d(t)~\hat{\cal G}_3/{\hg_2}, \cr
\hat\beta_2 &=\hg_3\dot{\bphi}_0 -2{\hat g}_2+(1-{\epsilon\over 2})\bar\Phi_0\hg_3+
\int_0^1dt~z^d(t)~\bigl( \hat{\cal G}_3^2-\hat{\cal G}_4\bigr), \cr
\hat\beta_3 &=\hg_4\dot{\bphi}_0-\bigl(1+{\epsilon\over 2}\bigr){\hat g}_3
+(1-{\epsilon\over 2})\bar\Phi_0\hg_4+
\int_0^1dt~z^d(t)~\bigl( -2\hat{\cal G}_3^3+3\hat{\cal G}_3\hat{\cal G}_4
-\hat{\cal G}_5\bigr),\cr
\hat\beta_4 &=\hg_5\dot{\bphi}_0-\epsilon {\hat g}_4+(1-{\epsilon\over 2})
\bar\Phi_0\hg_5+\int_0^1dt~z^d(t)~\bigl( 6\hat{\cal G}_3^4-12\hat{\cal G}_3^2
\hat{\cal G}_4+3\hat{\cal G}_4^2+4\hat{\cal G}_3\hat{\cal G}_5 
-\hat{\cal G}_6\bigr),\cdots}}
where $\hat{\cal G}_n={{\hat g}_n\over {z^2(t)+{\hat g}_2}}$. The absence of 
linear term implies $\bphi_0={2\over {(2-\epsilon){\hat g}_2}}\int_0^1
dt z^d(t){\hat{\cal G}}_3$. The flow of $\bphi_0$ is connected with 
that of $\hg_n$ with $n$ odd; naturally all odd coupling constants
vanish when $\bphi_0=0$. 

Consider the two smooth representations:
\eqn\zdef{ \rho_{k,b}(p)=e^{-a(p/k)^{b}} \longrightarrow
z(t)=\Bigl({-{{\rm ln}~t}\over a}\Bigr)^{1/b},}
and 
\eqn\zdm{ \rho_{k,m}(p)=[1+(p/k)^m]^{-1} \longrightarrow  z(t)=\Bigl({1\over t}
-1\Bigr)^{1/m}.} 
We choose $a={\rm ln}~2$ so that
$\rho_{k,b}(p=k)=1/2$ and the smoothness of the cutoff is controlled by a single
parameter $b$ \ref\jens. From Figures 5 and 6 depicted below, it is
clear that the sharper the cutoff the more rapid the convergence 
for $\nu$ in the polynomial truncation scheme. That is,
$m,b=\infty$ is the optimal value at which the minimum sensitivity condition
${d\nu\over dM}\vert_{b,m}=0$ is met. 
As $b(m)$ is decreased, $\nu$ oscillates 
with increasing amplitude. Comparing the two figures, one also notices 
a smaller oscillation amplitude for the exponential cutoff.

\medskip
\medskip

\centerline{\epsfbox{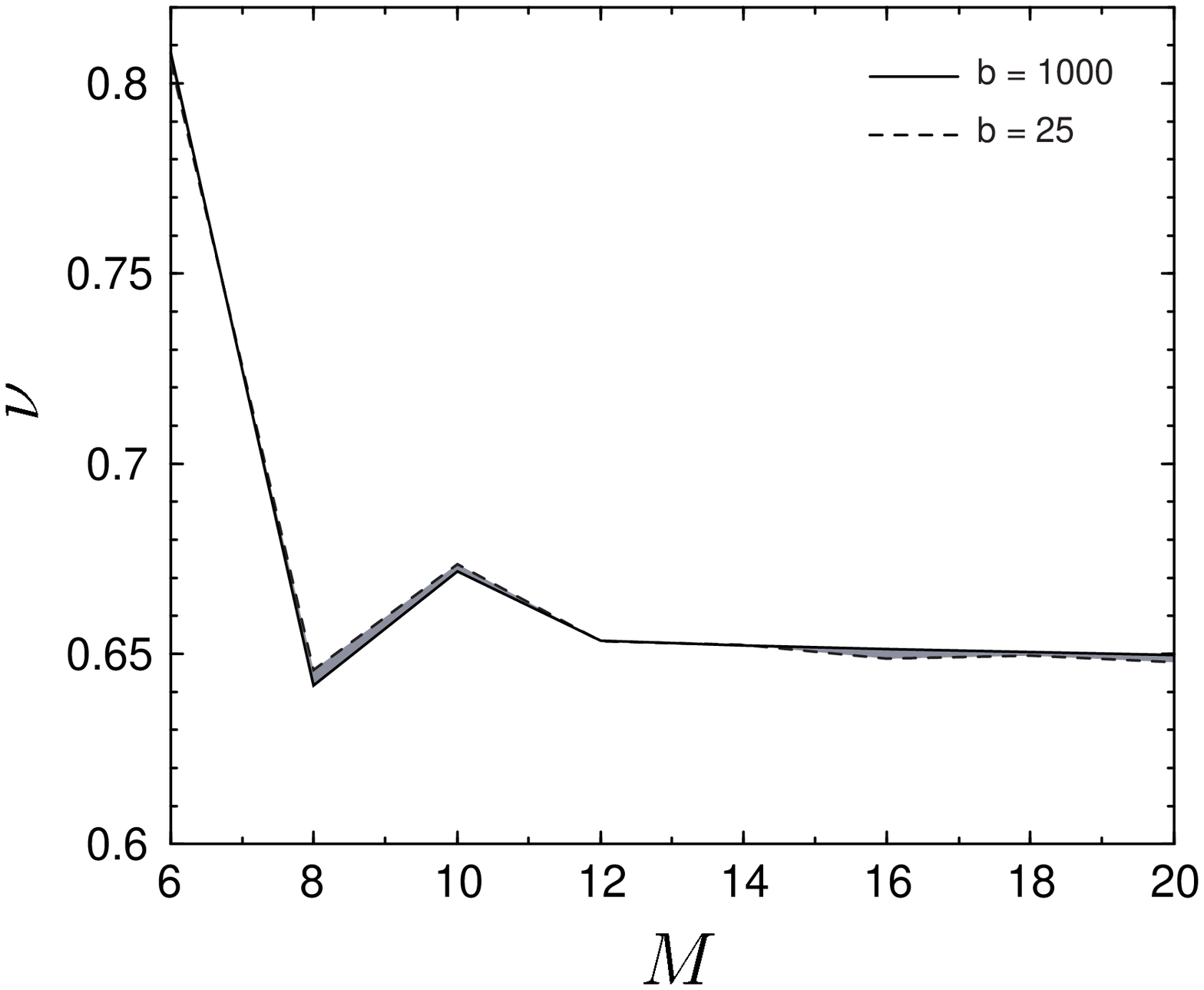}}
\medskip
{\narrower
{\sevenrm
{\baselineskip=8pt
\itemitem{Figure 5.}
Critical exponent $\scriptstyle \nu$ as a function of $\scriptstyle b$ for various
$\scriptstyle M$. The optimal smearing function corresponds to a sharp cutoff
with $\scriptstyle b=\infty$.
As $\scriptstyle b$ is decreased, $\scriptstyle \nu$ begins to oscillate 
with increasing amplitude. 
\bigskip
}}}

\medskip
\medskip

\centerline{\epsfbox{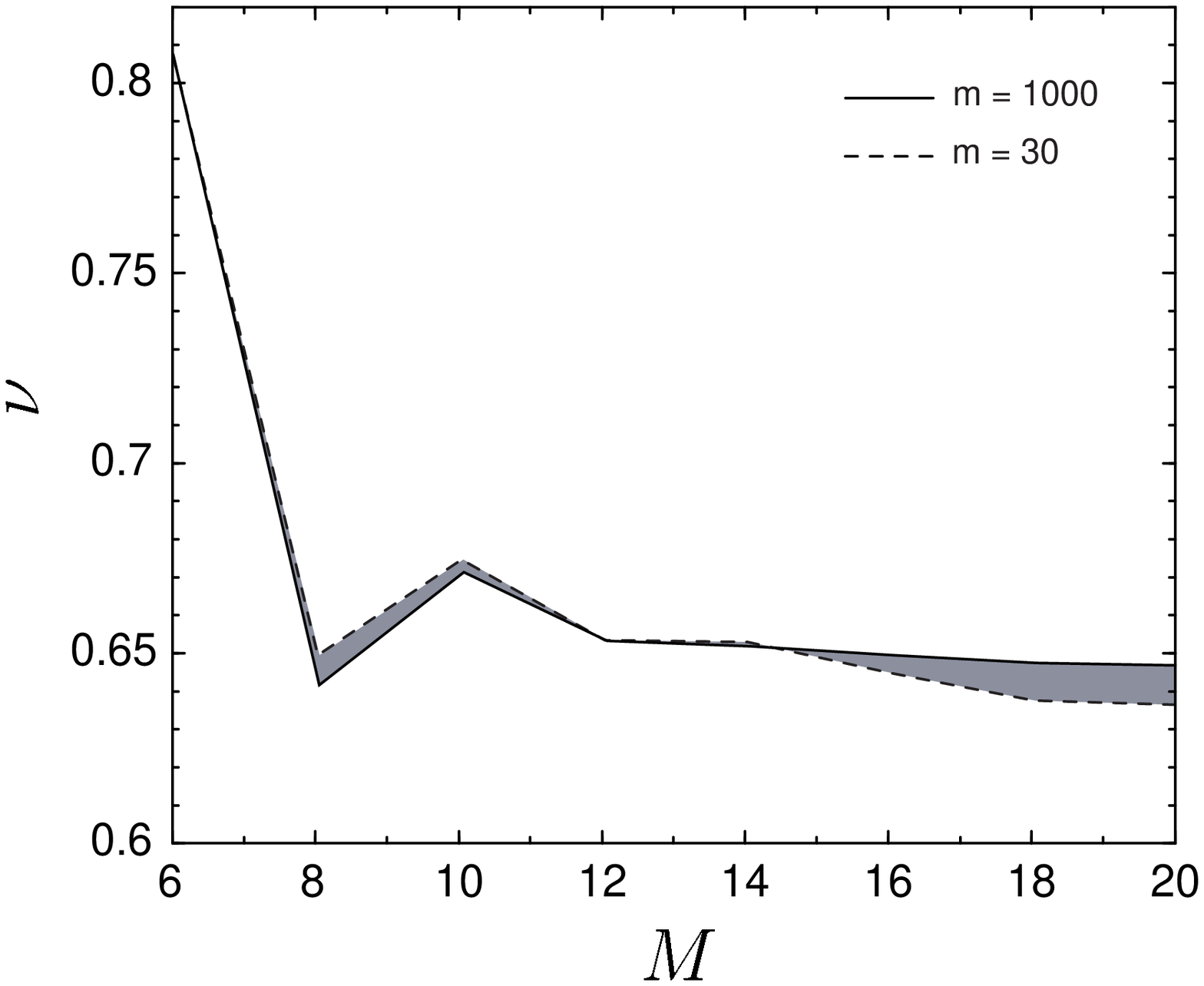}}
\medskip
{\narrower
{\sevenrm
{\baselineskip=8pt
\itemitem{Figure 6.}
Critical exponent $\scriptstyle \nu$ as a function of $\scriptstyle m$ for various
$\scriptstyle M$.
The ``optimal'' smoothness here again is $\scriptstyle m=\infty$.
\bigskip
}}}

\vskip 0.3in

We emphasize again that the optimal smoothness $\sigma$ 
obtained from the minimum sensitivity principle generally depends
on the model under study as well as the approximation employed. 
While a sharp cutoff is the optimal smoothness when $\bar\Phi'$ is chosen
to be the minimum $\bar\Phi_0$ of the potential, 
for an arbitrary $\bar\Phi'$, the optimal $\sigma$ which satisfies
the minimum sensitivity condition would correspond to a smooth cutoff.
On the other hand, a sharp cutoff yields a rather poor convergence for 
$\bar\Phi'=0$. A general criterion of finding
the optimal $\sigma$ can be found by analyzing the 
massless blocked propagator $\Delta_{\sigma}(z)=1/P^2_{\sigma}(z)
={\tilde\rho_{k,\sigma}/z^2}$, as has been recently shown by Litim 
(see Appendix B) \litim.

\medskip
\medskip
\centerline{\bf V. SUMMARY AND DISCUSSIONS}
\medskip
\nobreak
\xdef\secsym{5.}\global\meqno = 1
\medskip
\nobreak

In the present work we introduced a self-consistent RG prescription starting from
the one-loop perturbative expression.
Our self-consistent equation coincides with LPA in the sharp cutoff limit but
differs with a smooth cutoff. The advantage of this novel approach is that 
the perturbative independent-mode results can be readily ``dressed'' 
into a set of coupled non-linear RG equations. Complication arising from
taking the sharp cutoff limit in the derivative expansion can also be 
avoided completely. 

Due to the absence of a sharp boundary in the momentum 
integration in the smooth cases, 
a singularity is encountered in solving the  
truncated self-consistent equations, as the inverse propagator 
$z^2+\bg_2^{*}$ vanishes. We circumvent this difficulty 
by expanding ${\bar U}^{*}(\bar\Phi)$ 
about its minimum $\bar\Phi_0$ where $\hg_2^{*} > 0$.
Such expansion is non-invariant since it contains odd powers of $\bar\Phi$ that  
destroys $Z_2$ symmetry. 
The system remains, however, in the symmetric phase; as we
integrate down to $k=0$ in the dimensionful flow equation, 
the moving minimum $\Phi_0(k)$ vanishes. 
Nevertheless, the formal equivalence provides a useful tool for exploring systems
in the state of broken symmetry.  

In the absence of $Z_2$ symmetry, the physical, Wilson-Fisher fixed point which 
governs the critical behavior of the Ising universality class may be readily 
identified from the following conditions: (i) all the coupling constants
$\bg_n^{*}$'s are real; (ii) the potential has an extremum close to 
$\bphi_0^{*}=1.9287$, the value obtained without truncation; and (iii)  
the number of relevant eigen-directions is $j+1$ for $M=2j ~(j > 2)$.
Having identified the fixed point, the exponent $\nu$ is extracted 
from the truncated polynomial series. 
As shown in Sec. III, when a sharp cutoff is 
employed, $\nu$ oscillates with $M$ when ${\bar U}^{*}(\bar\Phi)$ is expanded
about $\bar\Phi=0$. On the other hand, the convergence improves dramatically
in the non-invariant expansion about the minimum $\bar\Phi_0$. We
truncated the series at $M=28$ where the eigenvalues become complex. 
From Figure 4, we see that with increasing $M$ the series converges rapidly to
approximately $\nu=0.649$, which differs from the full LPA solution $0.689$.
The discrepancy may be attributed to the difference in the symmetry of the two
approaches.
In Sec. IV where smooth cutoffs are considered, one finds that 
the amplitude of oscillation grows with a smaller $\sigma$, and 
the rate of convergence decreases. Thus, the sharp cutoff is the ``optimal''
smearing function for the self-consistent RG expanded about the minimum
$\bar\Phi_0$. Indeed, the minimal sensitivity condition 
${d\nu\over dM}\vert_{\sigma}=0$ is reached as $\sigma\to\infty$.

The results obtained in this paper demonstrates that 
when solving the truncated RG equations
for ${\bar U}_k(\bar\Phi)$, the convergence of physical quantities is 
strongly influenced by two factors: (i) the expansion point $\bar\Phi_0$, 
and (ii) the smoothness parameter $\sigma$ in the smearing 
function $\tilde\rho_{k,\sigma}(p)$. An optimal choice of $\bar\Phi_0$,
usually the minimum of ${\bar U}_k(\bar\Phi)$,
yields the largest radius of convergence in the expansion \aoki. However,
when $\bar\Phi_0$ is not optimal, one can fine-tune $\sigma$ so that 
physical quantities such as $\nu$ would exhibit minimal sensitivity on 
the order of truncation $M$. 

The optimization prescription presented here can be readily extended
to more complicated systems.
Polynomial truncation applied to models with fermionic interaction 
can be considered \ref\papp. A self-consistent RG prescription
with a smooth cutoff can also be formulated  
via the operator cutoff regularization in the proper-time space \wambach\ref\sbl.
In order to test the success of our self-consistent RG scheme, 
the effect of wave function 
renormalization constant $Z_k(\bar\Phi)$ must be incorporated. The results
will be reported in the forthcoming publication \ref\wave.

\bigskip
\goodbreak
\bigskip
\centerline{\bf ACKNOWLEDGMENTS}
\medskip
\nobreak
S.-B. L. is grateful to D. Litim, J. Polonyi and Y.-C. Tsai for valuable discussions
and the hospitality of the Institute for Nuclear Theory, University of Washington 
during his visit.
This work is supported in part by funds provided by the National Science 
Council of Taiwan under contract \#NSC-89-2112-M-194-005 and by the United States
Department of Energy Grant DG-FG03-97-ER41014. C-Y. Lin was partially supported by
FAPESP, Brazil.

\bigskip
\medskip
\medskip

\centerline{\bf APPENDIX A: SHARP RG FLOW BY POLYNOMIAL TRUNCATION}
\medskip
\nobreak \xdef\secsym{{\rm A}.}\global\meqno = 1
\medskip
\nobreak

In this Appendix, we provide some detail of the RG behavior near the 
Wilson-Fisher fixed point, based on the polynomial truncation.
The evolution equation obtained with a sharp cutoff reads, in $d=3$:
\eqn\gti{ {\dot{\bar U}}_k(\bphi)-{1\over 2}\bar\Phi{\bar U}_k'(\bphi)
+3{\bar U}_k(\bar\Phi)
=-{\rm ln}~\Bigl[1+{\bar U}_k''(\Phi)\Bigr].}

There are several ways to obtain an approximate solution to the fixed-point equation. 
We summarize two schemes below:
\medskip
\medskip
\item{1.} expansion about  $\bphi=0$:
\medskip

Here we assume the potential to be of the form:
\eqn\uex{ {\bar U}_k(\bar\Phi)=\sum_{n=1}^M{
{\bar g}_{2n}(k)\over{(2n)!}}~\bar\Phi^{2n},\qquad\qquad
{\bar g}_{2n}(k)={\bar U}^{(2n)}_k(0)={{\partial^{2n} {\bar U}_k}\over{\partial
\bar\Phi^{2n}}}\Big\vert_{\bar\Phi=0},}
and the flow is characterized by a set of $M$ coupled equations for ${\bar g}_{2n}$. 

\medskip

\item{(1)} $M=2$:
\medskip
At this order, we have
\eqn\snwo{\eqalign{\bar\beta_2 &=-2{\bar g}_2-{\bg_4\over{1+{\bar g}_2}},\cr
\bar\beta_4 &=-{\bar g}_4+{{3\bg_4^2}\over{(1+\bg_2)^2}}+O(\bg_6),}}
and the Wilson-Fisher fixed point is located at 
$({\bar g}_2^{*},{\bar g_4}^{*})= \bigl(-{1\over 7},~{12\over 49} \bigr)$.
Linearizing the flow then yields
\eqn\linef{ k{\partial\over{\partial k}}\pmatrix{{\bar g}_2\cr \cr {\bar
g}_4\cr}={\cal M}\pmatrix{{\bar g}_2\cr\cr {\bar g}_4\cr},}
where
\eqn\eem{{\cal M}=\pmatrix{ {{\partial{\bar\beta}_2}\over{\partial{\bar g}_2}} &
{{\partial{\bar\beta}_2}\over{\partial{\bar g}_4}} \cr \cr
{{\partial{\bar\beta}_4}\over{\partial{\bar g}_2}} &
{{\partial{\bar\beta}_4}\over{\partial{\bar g}_4}} \cr}_{{\bg}^*}
=\pmatrix{-2+{{\bar g}_4^{*}\over{(1+{\bar g}_2^{*})^2}}& -{1\over{1+{\bar
g}_2^{*}}}\cr \cr -{6{\bar g}_4^{2*}\over{(1+{\bar g}_2^{*})^3}} &
-\epsilon+{6{\bar g}_4^{*}\over{(1+{\bar g}_2^{*})^2}}\cr}
=\pmatrix{-2+{\epsilon\over 3}& -{6+\epsilon\over 6}\cr \cr
-{4\epsilon^2\over{6+\epsilon}} & \epsilon\cr}}
is the linearized RG matrix about the Wilson-Fisher fixed point.
Upon solving the eigenvalue problem for ${\cal M}$, we find
$\nu=0.527$ which deviates from the exact value $0.689$, found by solving the 
non-truncated equation \gti. 
The discrepancy is due to the fact that our approximate
Wilson-Fisher fixed point still lies too far away from the exact solution 
$(s_2, s_4)=(-0.4615, 0.4970)$. Thus in the neighborhood of our approximate solution, 
higher-order operators continue to evolve and therefore must be included in order 
to improve the accuracy. 

\medskip

\item{(2)} $M=3$:
\medskip

Taking into consideration the running of $\bg_6$, we have
\eqn\snwo{\eqalign{\bar\beta_2 &=-2{\bar g}_2-{\bg_4\over{1+{\bar g}_2}},\cr
\bar\beta_4 &=-{\bar g}_4+{{3\bg_4^2}\over{(1+\bg_2)^2}}-{\bg_6\over{1+\bg_2}},\cr
\bar\beta_6 &=-{{30\bg_4^3}\over{(1+\bg_2)^3}}
+{15\bg_4\bg_6\over{(1+\bg_2)^2}}+O(\bg_8),}}
which leads to
\eqn\wft{ \bigl({\bar g}_2^{*},{\bar g_4}^{*}, \bg_6^{*}\bigr)= \bigl(-{1\over 3},
~{4\over 9},~{16\over 27} \bigr).}
A comparison with the $M=2$ calculation shows that the fixed points are now closer
to the exact solution with the inclusion of $\bg_6$, 
and hence a more accurate $\nu$ is expected. Indeed for $M=3$
we find $\nu=0.585$. The complex singularities near the origin 
at this order are found to be
$\bphi_c=\pm 0.58998\pm 2.20183 i$, or $(|\bphi_c|,~\theta_c)=(2.2795,\pm 0.4167\pi)$.

As discussed in Sec. III, the terms from the $\epsilon$ expansion are reshuffled in
the polynomial truncation scheme. Retaining terms up to $O(\epsilon^3)$
at this order, we find
\eqn\gsim{ {\cal M}=\pmatrix{-2+{\epsilon\over 3}+{5\epsilon^2\over 9}
+{55\epsilon^3\over 54}& -\bigl(1+{\epsilon\over 6}+{5\epsilon^2\over 18}
+{55\epsilon^3\over 108}\bigr) & 0 \cr -{2\epsilon^2\over 3}-{14\epsilon^3\over
9} & \epsilon+{10\epsilon^2\over 3} +{55\epsilon^3\over 9} &
-\bigl(1+{\epsilon\over 6}+{5\epsilon^2\over 18} +{55\epsilon^3\over 108}\bigr)
\cr {10\epsilon^3\over 3} & -10\epsilon^2-{70\epsilon^3\over 3} &
2+3\epsilon+{25\epsilon^2\over 3}+{275\epsilon^3\over 18} \cr},}
which, for $\epsilon=1$, leads to 
\eqn\nulpa{ \nu={1\over 2}+{1\over 12}~\epsilon+{5\over 72}~\epsilon^2
+O(\epsilon^3),}
or $\nu\approx 0.653$. Comparing Eq.
\nulpa\ with that obtained in \zinn\
\eqn\nulpb{ \nu'={1\over 2}+{1\over 12}~\epsilon+{7\over 162}~\epsilon^2
+O(\epsilon^3),}
we see that they agree up to $O(\epsilon)$; the discrepancy in the $\epsilon^2$
term is due to the negligence of certain Feynman diagrams as well as
the wavefunction renormalization. 

\medskip

\item{(3)} $M \ge 4$:
\medskip

With increasing $M$, our approximate fixed point moves even closer to the 
non-truncated solution and a greater accuracy is attained. The results are summarized as 
follows:

\bigskip
\centerline{\sevenrm \vbox{\tabskip=0pt \offinterlineskip
\def\tablerule{\noalign{\hrule}}
\halign to375pt{\strut#&\vrule#\tabskip=1em plus2em&
  #\hfil& \vrule#& #\hfil& \vrule#& #\hfil& \vrule#& 
  #\hfil& \vrule#& #\hfil& \vrule#& #\hfil& \vrule#&
  #\hfil& \vrule#& #\hfil& \vrule#& #\hfil& \vrule# \tabskip=0pt\cr\tablerule
&& \omit\hidewidth $M$ 	                  \hidewidth
&& \omit\hidewidth $\bg_2^{*}$		  \hidewidth
&& \omit\hidewidth $\bg_4^{*}$  	  \hidewidth
&& \omit\hidewidth $\bg_6^{*}$		  \hidewidth
&& \omit\hidewidth $\bg_8^{*}$  	  \hidewidth
&& \omit\hidewidth $\bg_{10}^{*}$         \hidewidth
&& \omit\hidewidth $\bg_{12}^{*}$  	  \hidewidth
&& \omit\hidewidth $\bg_{14}^{*}$         \hidewidth
&& \omit\hidewidth $\bg_{16}^{*}$  	  \hidewidth &\cr\tablerule
&&  2 && -.1429 && .2449 &&   -   && - && - && - && - && -  &\cr\tablerule
&&  3 && -.3333 && .4444 && .5925 && - && - && - && - && -  &\cr\tablerule
&&  4 && -.4664 && .4977 && 1.1273 && 2.7804 && - && - && - && -  &\cr\tablerule
&&  5 && -.5163 && .4995 && 1.3056 && 4.2462 && 9.9214 && - && - && -  &\cr\tablerule
&&  6 && -.4508 && .4952 && 1.0675 && 2.3609 && -2.303 && -77.3326 && - && -  &\cr\tablerule
&&  7 && -.4349 && .4915 && 1.0048 && 1.9537 && -4.2865 && -84.4919 && -187.96 
&& -  &\cr\tablerule
&&  8 && -.4665 && .49776 && 1.1278 && 2.7837 && 0.0191 && -66.3291 && 248.552 
&& 18590.5  &\cr\tablerule
}}}

\bigskip
{\narrower
{\sevenrm
{\baselineskip=7pt
\centerline{Table 1. $\scriptstyle \bg_n^{*}$ as a function of $\scriptstyle M$.}
\bigskip
}}}

The relation between different $\bg^{*}_n$'s can be obtained by
expanding ${\bar U}^{*}$ about $\bar\Phi=0$:
\eqn\fpexp{ {\bar U}^{*}(\bphi)=-{1\over 3}~{\rm ln}(1+\bg^*_2)
+{\bg_2^*\over 2}~{\bar\Phi}^2
+\sum_{n=2}^{\infty}{\bg_{2n}^{*}\over (2n)!}~\bphi^{2n}.}
For large $M$, the value of $\bg_2^{*}$ approaches $s=-0.461533$
and $\bg_{2n}^{*}$ can be rewritten in terms of $s$ as:
\eqn\coef{\eqalign{ \bg^{*}_4 &=-2s(1+s), \cr
\bg_6^{*} &=2s(1+s)(1+7s), \cr
\bg_8^{*} &=-60 s^2(1+s)(1+3s), \cr
\bg_{10}^{*} &=40s^2(1+s)(2+43s+83s^2), \cr
\bg_{12}^{*} &=-80s^2(1+s)(-2+45s+549s^2+880s^3),\cdots}}
From the non-truncated RG equation we see that the potential 
becomes singular when the argument inside logarithm vanishes, i.e.,
$1+{\bar U}^{*''}(\bphi)=0$. 
Therefore, when expanding ${\rm ln}[1+{\bar U}^{*''}(\bphi)]$ in power 
series of $\bphi$, for large enough $M$ and $\bg_2^{*}\approx s$, one observes a 
four-fold periodic oscillation of $\bg_n^{*}$ in $(+,+,-,-)$ pattern,
due to the presence of a complex singularity near the origin
whose phase is approximately $\pi/2$.
Thus, increasing $M$ will at first improve the result by pushing the singularity
further away from the origin $\bar\Phi=0$ where the expansion is made, but 
eventually the series starts to oscillates about $\nu=0.689 \pm 0.008$.

\medskip
\bigskip
\item{2.} expansion about  $\bphi^2\ne 0$:
\medskip

In order to improve the convergence in polynomial truncation, 
Aoki {\it et al} has considered an expansion
about some non-vanishing field instead of the origin \aoki. 
Employing $\bar\chi=\bphi^2/2$ which preserves the
$Z_2$ symmetry of the original action, Eq. \drgti\ becomes
\eqn\drgt{ \Biggl[k{\partial\over{\partial k}}-\bigl(d-2)
\bar\chi{\partial\over{\partial\bar\chi}}+d\Biggr]{\bar U}_k(\bar\chi)
=-{\rm ln}~\Bigl[1+{\bar U}_k'(\chi)+2\bar\chi{\bar U}_k''(\chi)\Bigr].}
The expansion 
\eqn\ueo{ {\bar U}_k(\bar\chi)=\sum_{n=0}^{M}{
c_{n}(k)\over{n !}}~(\bar\chi-\bar\chi'(k))^{n},\qquad\qquad
c_{n}(k)={\bar U}^{(n)}_k(\chi')={{\partial^{n} {\bar U}_k}\over{\partial
\bar\chi^{n}}}\Big\vert_{\bar\chi=\bar\chi'},}
leads to 
\eqn\nwy{\eqalign{ {\dot c}_0 &={\dot\chi}'c_1-(4-\epsilon)c_0+(2-\epsilon)\chi'c_1
-{\rm ln}\bigl(1+c_1+2\chi'c_2\bigr), \cr
{\dot c}_1 &= {\dot\chi}'c_2 -2c_1-\chi'\bigl[(-2+\epsilon)c_2+2{\hat c}_3\bigr]
-3{\hat c}_2, \cr
{\dot c}_2 &= {\dot\chi}'c_3 -(4-\epsilon)c_2+(2-\epsilon)(2c_2+\chi'c_3)+(3{\hat c}_2
+2\chi'{\hat c}_3)^2-(5{\hat c}_3+2\chi'{\hat c}_4), \cr
{\dot c}_3 &= {\dot\chi}'c_4 -(4-\epsilon)c_3+(2-\epsilon)(3c_3+\chi'c_4)
-2(3{\hat c}_2+2\chi'{\hat c}_3)^3,\cr
&\quad
+3(3{\hat c}_2+2\chi'{\hat c}_3)(5{\hat c}_3+2\chi'{\hat c}_4)
-7{\hat c}_4 +O(c^5),\cr
{\dot c}_4 &= (4-3\epsilon)c_4+6(3{\hat c}_2+2\chi'{\hat c}_3)^4
+28(3{\hat c}_2+2\chi'{\hat c}_3){\hat c}_4
-12(3{\hat c}_2+2\chi'{\hat c}_3)^2(5{\hat c}_3+2\chi'{\hat c}_4),\cr
&\quad
+3(5{\hat c}_3+2\chi'{\hat c}_4)^2+O(c^5,c^6),\cdots}}
where ${\hat c}_n=c_n/(1+c_1+2\chi'c_2)$. 
As illustrated in Sec. III, a more rapid convergence would be
obtained if the expansion is made around ore near the minimum $\chi_0=1.844$.
We summarize in Table 2 below for the results obtained about $\chi'=2$:
From the Table we see a remarkable convergence of $\nu$ toward the full LPA
solution in this expansion scheme.  

\bigskip
\centerline{\sevenrm \vbox{\tabskip=0pt \offinterlineskip
\def\tablerule{\noalign{\hrule}}
\halign to400pt{\strut#&\vrule#\tabskip=.5em plus .5em&
  #\hfil& \vrule#& #\hfil& \vrule#& #\hfil& \vrule#& #\hfil& \vrule#& 
  #\hfil& \vrule#& #\hfil& \vrule#& #\hfil& \vrule#& #\hfil& \vrule#&
  #\hfil& \vrule#& #\hfil& \vrule#& #\hfil& \vrule# \tabskip=0pt\cr\tablerule
&& \omit\hidewidth $M$ 	                  \hidewidth
&& \omit\hidewidth $\nu$ 	           \hidewidth
&& \omit\hidewidth $c_1$		  \hidewidth
&& \omit\hidewidth $c_2$         	  \hidewidth
&& \omit\hidewidth $c_3$		  \hidewidth
&& \omit\hidewidth $c_4$  	           \hidewidth
&& \omit\hidewidth $c_5$                   \hidewidth
&& \omit\hidewidth $c_6$  	            \hidewidth
&& \omit\hidewidth $c_7$                    \hidewidth
&& \omit\hidewidth $c_8$                    \hidewidth
&& \omit\hidewidth $c_9$  	             \hidewidth &\cr\tablerule
&&  3 && .66781 && -.0084 && .2664 && .0829 && - && - && - && - && - &&-&\cr\tablerule
&&  4 && .68698 &&  .0303 && .3343 && .1091 &&.0137 && - && - && - &&- &&-&\cr\tablerule
&&  5 && .68803 &&  .0490 && .3586 && .1155 &&.0145 &&-.0071 && - && -&&-&&-  &\cr\tablerule
&&  6 && .68907 && .0512 && .3619 && .1167 &&.0151 &&-.0072 && .0016 &&-&&-&&-&\cr\tablerule
&&  7 && .68956 &&  .0498 && .3601 && .1162 && .0150 && -.0066 && .0016
&& .0026  &&-&&-&\cr\tablerule
&&  8 && .68952 && .0492 && .3591 && .1159 && .0148 && -.0067 && .0011 
&& .0023 && -.0040&&-  &\cr\tablerule
&&  9 && .68948 && .0491 && .3590 && .1159 && .0148 && -.0066 && .0011 
&& .0025 && -.0040&&.0012  &\cr\tablerule
}}}

\bigskip
{\narrower
{\sevenrm
{\baselineskip=7pt
\centerline{Table 2. The critical exponent $\scriptstyle \nu$ coupling constants 
$\scriptstyle c_n$ as a function of $\scriptstyle M$.}
\bigskip
}}}

\medskip
\medskip

\centerline{\epsfbox{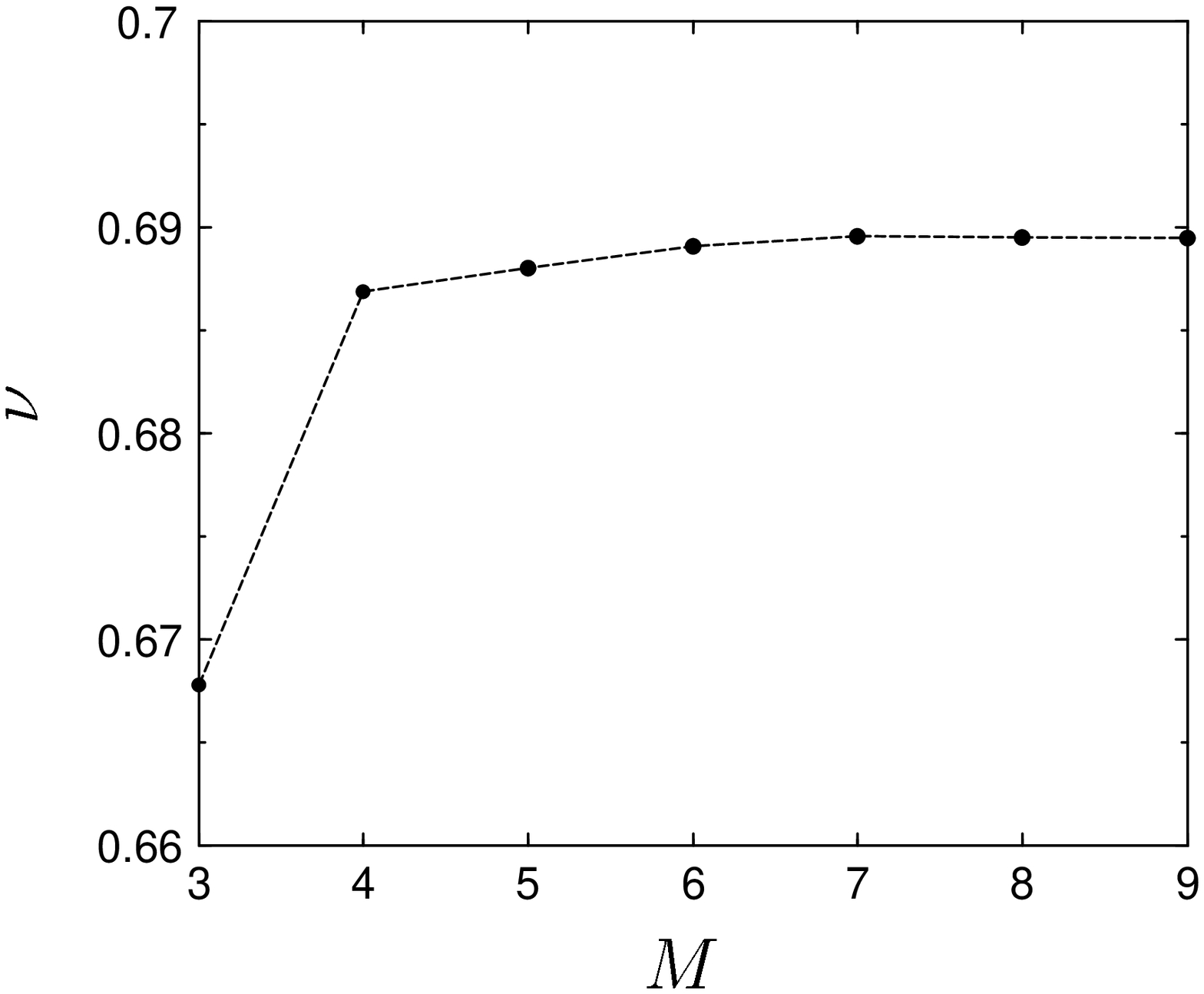}}
\medskip
{\narrower
{\sevenrm
{\baselineskip=8pt
\itemitem{Figure 7.}
Critical exponent $\scriptstyle \nu$ as a function of $\scriptstyle M$ 
obtained using the invariant expansion about $\scriptstyle \chi'=2$.
\bigskip
\bigskip
}}}


\medskip
\medskip
\centerline{\bf APPENDIX B: MODIFIED BLOCKED PROPAGATORS}
\medskip
\nobreak \xdef\secsym{{\rm B}.}\global\meqno = 1
\medskip
\nobreak

We have seen that with a smooth cutoff the momentum integrations for the 
fixed-point solution are divergent in the self-consistent RG formalism,
while they remain completely analytic in the LPA. The distinction can be readily
understood from the manner in which the propagators are modified in the two approaches. 
\medskip
\medskip
\centerline{\bf a. exact RG}
\medskip
\medskip

In the exact RG approach, the second functional determinant is given by 
${\widetilde S}_k''[\Phi]=C_k^{-1}+S''[\Phi]$ where 
$C_k^{-1}=p^2(1-\tilde\rho_{k,\sigma})/{\tilde\rho}_{k,\sigma}$. This in turn implies
a modified (dimensionless) propagator of the form
\eqn\mop{\Delta_{\sigma}(z)={1\over{ P^2_{\sigma}(z)+\bg_2}},}
where $z=p/k$, $P^2_{\sigma}(z)=z^2/{\tilde\rho_{k,\sigma}}$ and $\bg_2
={\bar U}''_k(0)$.
Let us first examine the behavior of $P^2_{\sigma}(z)$. 
For definiteness we take the smooth smearing functions to be $\tilde\rho_{b}(z)
=1-e^{-az^b}$ as considered in Sec. IV. 

We see that as $z$ is decreased, with the exception of $b=2$, 
$P^2_b$ attains a minimum and then diverges as
$z\to 0$. The minimum $z_0$, located by ${\partial P^2_b}/{\partial z}=0$, 
satisfies the relation $z_0^be^{-az_0^b}/(1-e^{-az_0^b})=2/ab$ and implies a minimum 
$P^2(z_0)=2z_0^{2-b}e^{az_0^b}/ab$. As can be seen from Table 3,
$P^2_b(z_0)$ approaches unity in the limit $b\to\infty$. 
As recently addressed by Litim \litim, $b$ can be chosen in such a way as 
to achieve the maximum radius of 
convergence $R=\lim_{n\to\infty}a_n/a_{n+2}$
in the amplitude expansion whose coefficients are given by
\eqn\ampl{ a_n=-\int_0^{\infty}dz~z^{d+1}\Bigl(k{\partial{\tilde\rho_{k,\sigma}(z)}
\over{\partial k}}\Bigr)P_{\sigma}^{-n}(z).}
By using the following criterion \litim:
\eqn\optim{ R_{\rm opt}={\rm max}\biggl({\rm min}_{z \ge 0}
~P^2_{\sigma}(z)\biggr),}
the optimal smoothness parameter associated with this particular class of 
smearing functions is given by $b_{\rm opt}=2/{\rm ln}2$, with $R_{\rm opt}=2$.  
On the other hand, based on the principle of minimum sensitivity, it was 
found that $b^{*}=3$ yields the fastest rate of convergence 
with respct to polynomial expansion for $\nu$ \lps. 
We emphasize that the 
two optimization procedures are inherently different; while $b_{\rm opt}$ 
is a universal result for the exponential smearing function, the vale of
$b^{*}$ depends on the model and the physical quantity under study \ref\dlitim.   

\medskip
\medskip

$$\hbox{\vbox{\offinterlineskip
\def\strut{\hbox{\vrule height 9pt depth 6pt width 0pt}}
\hrule
\halign{
\strut\vrule#\tabskip 0.1in&
\hfil$#$\hfil& \vrule#&   \hfil$#$\hfil& \vrule#& \hfil$#$\hfil&
\vrule#\tabskip 0.0in\cr
& b    && z_0     && P^2_b(z_0) & \cr\noalign{\hrule}
& 2    && 0       && 1/{\rm ln}2=1.4427 & \cr\noalign{\hrule}
& 2/{\rm ln}2 && 1   &&   2    & \cr\noalign{\hrule}
& 3    && 1.0324  && 1.9974    & \cr\noalign{\hrule}
& 4    && 1.1603  && 1.8821    & \cr\noalign{\hrule}
& 5    && 1.1849  && 1.7508    & \cr\noalign{\hrule}
& 20   && 1.0861  && 1.2122    & \cr\noalign{\hrule}
}}}$$

\bigskip
{\narrower
{\sevenrm
{\baselineskip=7pt
\centerline{Table 3. Location of the minimum of $\scriptstyle P^2_b$ as a function of 
$\scriptstyle b$.}
\bigskip
}}}
One may also consider the power-law cutoff
$\tilde\rho_m(z)=z^m/(1+z^m)$. 
From Table 4 we may readily conclude that $m_{\rm opt}=4$. The minimum 
sensitivity condition utilized in Ref. \lps\ gives $m^{*}=5.5$.

\medskip
$$\hbox{\vbox{\offinterlineskip
\def\strut{\hbox{\vrule height 9pt depth 6pt width 0pt}}
\hrule
\halign{
\strut\vrule#\tabskip 0.08in&
\hfil$#$\hfil& \vrule#&   \hfil$#$\hfil& \vrule#& \hfil$#$\hfil&
\vrule#\tabskip 0.0in\cr
& m && z_0=({{m-2}\over 2})^{1/m} && P^2_m(z_0)=({{m-2}\over 2})^{2/m}({m\over{m-2}})
&\cr\noalign{\hrule}
& 2    && 0       && 1  & \cr\noalign{\hrule}
& 3    && .7937  && 1.8899    & \cr\noalign{\hrule}
& 4    && 1  && 2    & \cr\noalign{\hrule}
& 5.5    && 1.1071  && 1.9261    & \cr\noalign{\hrule}
& 20   && 1.1161  && 1.3841    & \cr\noalign{\hrule}
}}}$$

\bigskip
{\narrower
{\sevenrm
{\baselineskip=7pt
\centerline{Table 4. Location of the minimum of $\scriptstyle P^2_m$ as a function of 
$\scriptstyle m$.}
\bigskip
}}}

In Fig. 8 the blocked propagator is depicted for the exponential cutoff.
Note that for $b=2$, $1/P^2_b(z)$ acquires an effective mass which remains finite
in the limit $z\to 0$.

\medskip
\medskip

\centerline{\epsfbox{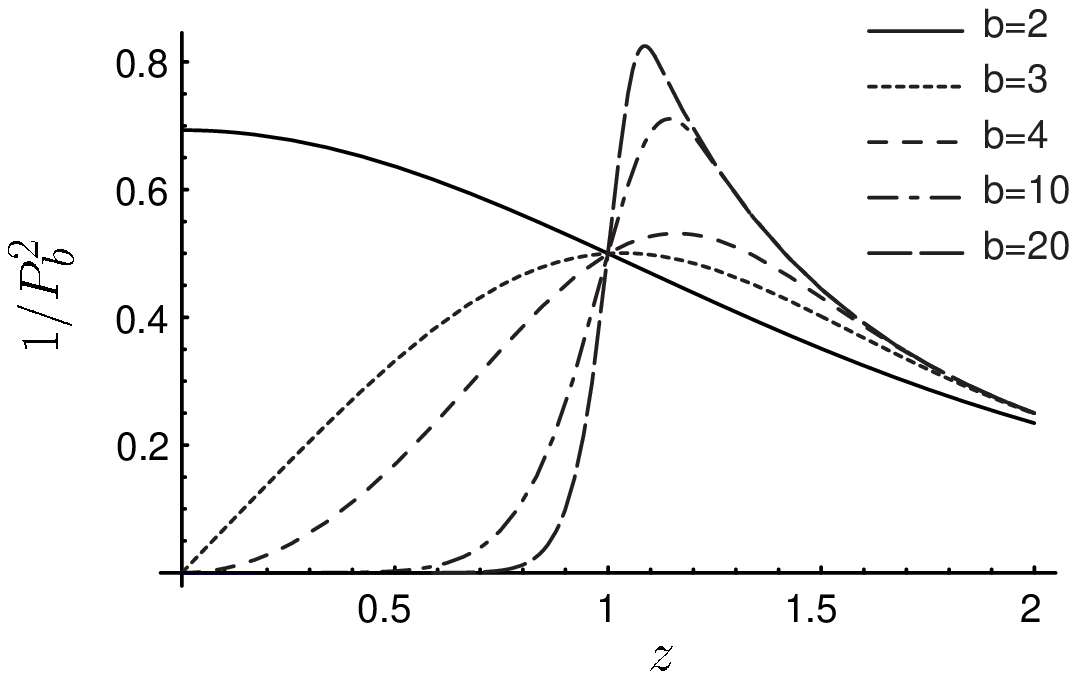}}
\medskip
{\narrower
{\sevenrm
{\baselineskip=8pt
\itemitem{Figure 8.}
$\scriptstyle 1/P^2_b(z)$ as a function of 
$\scriptstyle z$ with $\scriptstyle a={\rm ln}2$
and $\scriptstyle b=2,3,4,10,20$. The propagators is IR finite
as $\scriptstyle z\to 0$. 
\bigskip
\bigskip
}}}

\vskip 0.15in

Let us now consider the effect of $\bg_2$. In the symmetric phase where
$\bg_2 > 0$, the change is only a shift $P^2_b\to P^2_b+\bg_2$. On the other hand,
if $\bg_2 < 0$ as in the case of spontaneous symmetry breaking or during the course of RG
evolution where $\bg_2$ is still negative, the modified propagator can become singular
if $P^2_b(z)+\bg_2 \le 0$. However, this does not happen in LPA where 
$-\bg_2^{*}=0.461533 < {\rm min} P_b^2(z)=1$. Thus in 
the LPA
prescription with a smooth cutoff, the momentum integrations are completely analytic.
We comment that in the case of spontaneous symmetry breaking where $\bg_2$ remains 
negative down to $k=0$, one expects the inverse (dimensionful) propagator 
$(p^2/{\tilde\rho}_b)+g_2$ to 
vanish at some scale $k_{\rm cr}$, and the region below $k < k_{\rm cr}$ 
becomes inaccessible by conventional perturbation theory.
\medskip
\medskip
\centerline{\bf b. self-consistent RG}
\medskip
\medskip

Let us now turn to the self-consistent RG prescription. In this formalism, the 
modified propagator reads
\eqn\mopp{\Delta_{\sigma}(z)={{\tilde\rho}_{\sigma}(z)\over{ z^2+\bg_2}}.}
Contrary to Eq. \mop, $\bg_2$ is also influenced by the presence of 
${\tilde\rho}_{\sigma}(z)$ and must not be treated separately. 
Using the exponential smearing function $\tilde\rho_b(z)$ as an example, 
we find that for $\bg_2 > 0$ 
the qualitative feature of the modified propagator remains the same except for 
$b=2$; while $P^2_b$ approaches $1/{\rm ln}2$ as $z\to 0$ in the exact RG formalism, 
it diverges
in the self-consistent RG prescription. The minimum of ${\tilde P}^2_b(z)$ is located 
by solving 
\eqn\minz{ {{z_0^2+\bg_2}\over{1-e^{-az_0^b}}}={2z_0^{2-b}e^{az_0^b}\over ab}.}

The major difference between the two
schemes occurs when $\bg_2 < 0$. As shown in Fig. 9 below,  
while the propagator in the former remains analytic (for $\bg_2 > -1$), 
singularity is present in the latter. Thus, when employing the self-consistent RG 
prescription to analyze the fixed-point solution, it is desirable to 
expand about some $\bar\Phi_0$ where ${\bar U}''_k(\bar\Phi_0) > 0$.

\medskip
\medskip
\medskip

\centerline{\epsfbox{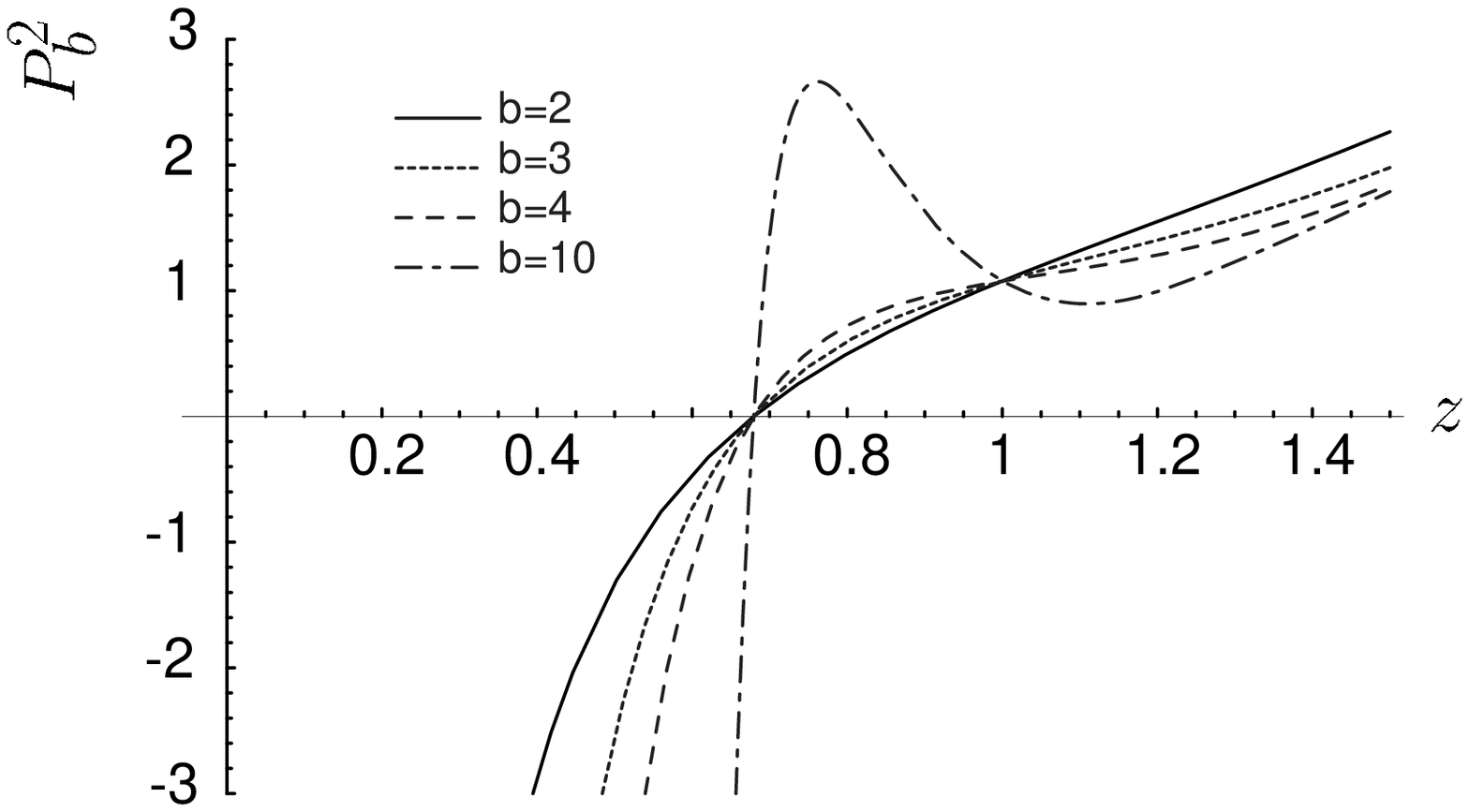}}
\medskip
\medskip
{\narrower
{\sevenrm
{\baselineskip=8pt
\itemitem{Figure 9.}
$\scriptstyle {\tilde P}^2_b(z)$ as a function of $\scriptstyle z$ with 
$\scriptstyle a={\rm ln}2$
and $\scriptstyle b=2,3,4$ and $\scriptstyle 10$ for $\scriptstyle \bg_2=-0.461533 < 0$.
The inverse propagator becomes singular when $\scriptstyle z \le \sqrt{-\bg_2^{*}}$. 
\bigskip
}}}

\vskip 0.15in

\medskip
\medskip

\centerline{\bf APPENDIX C: ASYMPTOTIC EXPANSION WITH SMOOTH SMEARING}
\medskip
\nobreak \xdef\secsym{{\rm C}.}\global\meqno = 1
\medskip
\nobreak

In Sec. IV we see that the momentum integrations for the fixed-point solution
in the self-consistent RG are divergent due to the negative fixed-point value $\bg_2^{*}$. 
Since $\bg_2^{*}\approx-0.4615$ is a ``small'' negative number, we inquire how good
the approximation is to make an asymptotic expansion about $\bg_2^{*}=0$. 

To explore the asymptotic limit, we first expand $U^{*}(\bphi)$ about the origin,
and the $\bar\beta$ functions read
\eqn\betafu{\eqalign{\bar\beta_2 &=-2{\bar g}_2-{\bar g}_4 I_1 \cr \bar\beta_4
&=-\epsilon {\bar g}_4+3{\bar g}_4^2 I_2-{\bar g}_6 I_1 \cr \bar\beta_6
&=(2d-6){\bar g}_6-30{\bar g}_4^3 I_3+15{\bar g}_4{\bar g}_6 I_2-{\bar g}_8
I_1,\cdots}}
where
\eqn\ibge{ I_n({\bar g}_2)=\int_0^1 dt~{z^d(t)\over{\bigl[z^2(t)+{\bar
g}_2\bigr]^n}}=\int_0^1{{dzz^{d-1}(abz^be^{-az^b})}\over
{\bigl(z^2+{\bar g}_2\bigr)^n}}= -\int_0^1{dzz^{d-1}\over
\bigl(z^2+{\bar g}_2\bigr)^n} ~\Bigl(k{{\partial\tilde\rho_{k,b}}
\over{\partial k}}\Bigr).}
At $M=2$ the $2\times 2$ linearized RG matrix becomes
\eqn\lineg{ {\cal M}_b=\pmatrix{-2+{\bar g}_4^{*}I_2({\bar g}_2^{*}) & -I_1({\bar
g}_2^{*})\cr \cr -6{\bar g}_4^{2*}I_3({\bar g}_2^{*}) & -\epsilon +6{\bar
g}_4^{*}I_2({\bar g}_2^{*})\cr},}
where the non-trivial Wilson-Fisher fixed points are located by solving the
transcendental equations
\eqn\wffp{ 0=6{\bar g}_2^{*}I_2({\bar g}_2^{*})+\epsilon I_1({\bar g}_2^{*})
=\int_0^1dt~{{z^d(t)\bigl[(6+\epsilon){\bar g}_2^{*}+\epsilon
z^2(t)\bigr]}\over {\bigl(z^2(t)+{\bar g}_2^{*}\bigr)^2}},}
and ${\bar g}_4^{*}={\epsilon/{3I_2({\bar g}_2^{*})}}$.
Note that in the limit $b\to\infty$, $I_n\to (1+{\bar g}_2)^{-n}$, and we
readily recover the sharp cutoff results.

Expanding $I_n(\bg_2)$ about ${\bar g}_2=0$:
\eqn\rgyi{\eqalign{I_n({\bar g}_2) &=ab \int_0^{\infty}{dz z^{d-1+b-2n}
e^{-az^b} \over {\bigl[1+\bigl({\bar g}_2/z^2\bigr)\bigr]^n}} \cr
& 
=\sum_{i=0}^{\infty}(-{\bar g}_2)^i {i+n-1\choose{n-1}}
a^{-{d-2(i+n)\over b}}~\Gamma\Bigl({b+d-2(i+n)\over b}\Bigr),}}
we have for $\epsilon=1$, 
\eqn\wfbe{ ({\bar g}_2^{*},{\bar g}_4^{*})= \biggl(
{{5c_1-\sqrt{25c_1^2+44c_0c_2}}\over 22c_2}, {{6c_1-\sqrt{25c_1^2+44c_0c_2}}\over
3(c_1^2-4c_0c_2)}\biggr),}
and
\eqn\lineg{ {\cal M}_b =\pmatrix{-{5\over 3} & {{3(5c_1^2-44c_0c_2-c1\sqrt{25c_1^2
+44c_0c_2})}\over 121c_2^2} \cr\cr
-{2c_2(-6c_1+\sqrt{25c_1^2+44c_0c_2})^2\over {3(c_1^2-4c_0c_2)^2}} & 1\cr},}
where $c_0=a^{-1/b}\Gamma(1+1/b)$, $c_1=a^{1/b}\Gamma(1-1/b)$, 
and $c_2=\Gamma(1-3/b)$.
The $b$ dependence of the exponent $\nu$ is depicted in
Fig. 10 below.
From the Figure, we see that the expansion is only reliable for sufficiently large $b$.
Indeed as $b\to\infty$, $\nu$ approaches $0.525$ which is the sharp cutoff result for $M=2$.

\medskip
\medskip

\centerline{\epsfbox{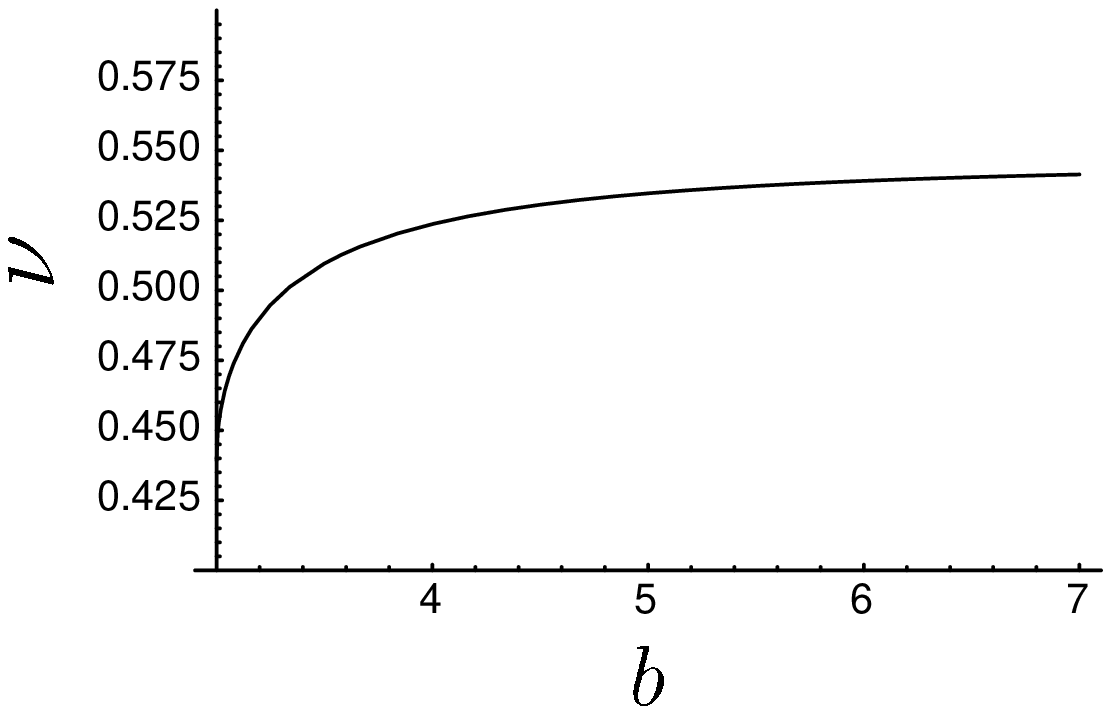}}
\medskip
{\narrower
{\sevenrm
{\baselineskip=8pt
\centerline{{Figure 10.}
Asymptotic value of $\scriptstyle \nu$ as a function of $\scriptstyle b$.}
\bigskip
}}}

\vskip 0.15in 

The asymptotic limit with $\tilde\rho_{k,m}={(p/k)^m\over {1+(p/k)^m}}$ can be explored 
in a similar manner. In this case we have
\eqn\ibge{ I_n({\bar g}_2)= -\int_0^{\infty}{dzz^{d-1}\over
\bigl(z^2+{\bar g}_2\bigr)^n} ~\Bigl(k{{\partial\tilde\rho_{k,m}}
\over{\partial k}}\Bigr)
=m\int_0^{\infty}{{dzz^{d+m-1}}\over
{\bigl(z^2+{\bar g}_2\bigr)^n(1+z^m)^2}},}
and the fixed-point solution for $M=2$ is found by solving
\eqn\wffpm{ 0=6{\bar g}_2^{*}I_2({\bar g}_2^{*})+\epsilon I_1({\bar g}_2^{*})
=m\int_0^{\infty}{{dz~z^{m+2}\bigl(7{\bar g}_2^{*}+z^2\bigr)}\over
{(1+z^m)^2(z^2+{\bar g}^{*}_2)^2}}.}
Making an asymptotic expansion, we have
\eqn\inap{I_n({\bar g}_2)=\sum_{i=0}^{\infty}(-{\bar g}_2)^i {i+n-1\choose{n-1}}
{{\pi(d-2(i+n))}\over m}~{\rm csc}\Bigl({{\pi(d-2(i+n))}\over m}\Bigr),}
provided that $d-2(i+n)+m > 0$. The resulting linearized RG matrix takes on the 
same form as Eq. \lineg, where $c_0=c_1={\pi\over m}{\rm csc}({\pi\over m})$ and  
$c_2=3{\pi\over m}{\rm csc}({3\pi\over m})$.
The $m$ dependence of $\nu$ is shown in Fig. 11.
Once more, the results are only reliable in the large $m$ limit,
where $\nu \approx 0.525$, in agreement with that obtained using the sharp cutoff.

\medskip
\medskip

\centerline{\epsfbox{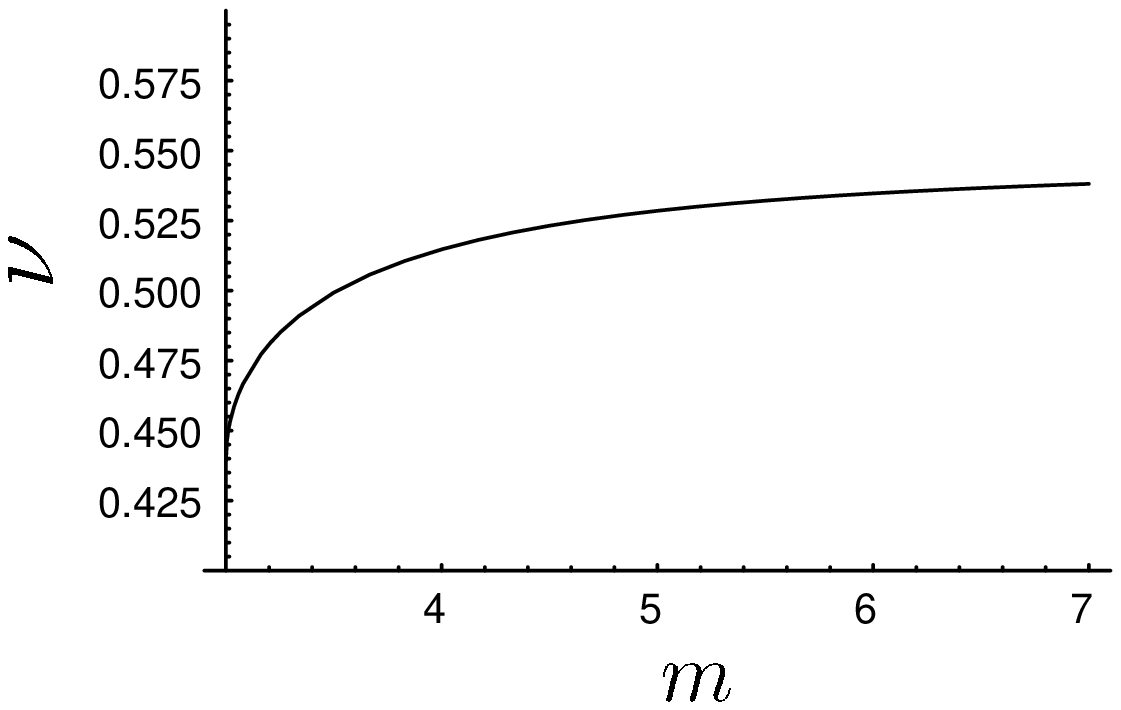}}
\medskip
{\narrower
{\sevenrm
{\baselineskip=8pt
\centerline{{Figure 11.}
Critical exponent $\scriptstyle \nu$ as a function of $\scriptstyle m$.}
\bigskip
}}}

\vskip 0.3in

\bigskip
\medskip

\medskip
\centerline{\bf REFERENCES}
\medskip
\medskip
\nobreak

\item{\wilson} K. Wilson, {\it Phys. Rev.} {\bf B4} (1971) 3174;
K. Wilson and J. Kogut, {\it Phys. Rep.} {\bf 12C} (1975) 75.
\medskip
\item{\wegner} F.J. Wegner and A. Houghton, {\it Phys. Rev}
{\bf A8} (1972) 401.
\medskip
\item{\polchinski} J. Polchinski, {\it Nucl. Phys.}
{\bf B231} (1984) 269.
\medskip
\item{\morris} T. R. Morris,
{\it Phys. Lett.} {\bf B334} (1994) 355 and {\bf B329} (1994) 241;
\medskip
\item{\wetterich} see, for example,  J. Berges, N. Tetradis and C. Wetterich, 
hep-ph/0005122 and references therein.
\medskip
N. Tetradis and C. Wetterich, {\it Nucl. Phys.} {\bf B422} (1994) 541;
\medskip
B. Bergerhoff and J. Reingruber, {\it Phys. Rev.} {\bf D60} (1999) 105036;
\medskip
B. Bergerhoff, J. Manus, J. Reingruber,  {\it ibid.} {\bf D61} (2000) 125005; 
\medskip
B. Bergerhoff, {\it Phys. Lett.} {\bf B437} (1998) 381;
\medskip
\item{\others}
C. Bagnuls and C. Bervillier, hep-th/0002034. 
\medskip
J. Comellas and A. Travesset, {\it Nucl/ Phys.} {\bf B498} (1997) 539.
\medskip
\item{\tmorris} T. R. Morris, {\it Int. J. Mod. Phys.} {\bf A9} (1994) 2411;
{\it Nucl. Phys.} {\bf B509} (1998) 637 and {\bf B495} (1997) 477, and references therein.
\medskip
\item{\sb} S.-B. Liao and J. Polonyi, {\it Ann. Phys.} {\bf 222} (1993) 122
and {\it Phys. Rev.} {\bf D51} (1995) 4474.
\medskip
\item{\nicoll} J. F. Nicoll, T. S. Chang and H. E. Stanley, {\it Phys. Rev.
Lett.} {\bf 32} (1974) 1446, {\bf 33} (1974) 540; {\it Phys. Rev.} {\bf A13}
(1976) 1251.
\medskip
\item{\mike} S.-B. Liao and M. Strickland, {\it Nucl. Phys.} {\bf B497}
(1997) 611 and {\it ibid.} {\bf B532} (1998) 753. 
\medskip
\item{\hasenfratz} A. Hasenfratz and P. Hasenfratz, {\it Nucl.}
{\bf B270} (1986) 687.
\medskip
\item{\margaritis} A. Margaritis, G. Odor and A. Patkos, {\it Z. Phys.}
{\bf C39} (1988) 109.
\medskip
\item{\aoki} K. Aoki, K. Morikawa, W. Souma, J. Sumi and H. Terao,
{\it Prog. Theor. Phys.} {\bf 95} (1996) 409 and {\bf 99} (1998) 451.
\medskip
\item{\lps} S.-B. Liao, J. Polonyi  and M. Strickland, {\it Nucl. Phys.} {\bf B567}
(2000) 493.
\medskip
\item{\litim} D. Litim, {\it Phys. Lett.} {\bf B486} (2000) 92.
\medskip
\item{\lp} S.-B. Liao, J. Polonyi and D. P. Xu, {\it Phys. Rev.}
{\bf D51} (1995) 748;
\medskip 
S.-B. Liao and M. Strickland, {\it ibid.} {\bf D52} (1995) 3653.
\medskip
\item{\alexandre}J. Alexandre and J. Polonyi, hep-th/9902144.  
\medskip
\item{\enzo}  V. Branchina, {\it Phys. Rev.} {\bf D62} (2000) 065010;
\medskip 
A. Bonanno, V. Branchina, H.Mohrbach and D. Zappala, {\it ibid.} {\bf D60} (1999) 065009;
\medskip
A. Bonanno and D. Zappala, {\it ibid.} {\bf D57} (1998) 7383. 
\medskip
\item{\mthesis} M. Strickland, Ph.D. thesis and hep-ph/9809592. 
\medskip
\item{\chiral} see, for example, K. Aoki, K. Takagi, H. Terao and M. Tomoyose, 
hep-th/0002038 and references therein.
\medskip
\item{\wambach} O. Bohr, B.-J. Schaefer and J. Wambach, hep-ph/0007098.
\medskip
\item{\jean} J. Alexandre, V. Branchina and J. Polonyi, {\it Phys. Rev.} {\bf D58} (1998) 
016002.
\medskip
\item{\zinn} J. Zinn-Justin, {\it Quantum Field Theory and Critical
Phenomena}, 3rd ed. (Clarendon, Oxford, 1996).
\medskip
\item{\jens} J. O. Andersen and M. Strickland, cond-mat/9808346 and cond-mat/9811096.
\medskip
\item{\papp} G. Papp, B.-J. Schaefer, H.-J. Pirner and J. Wambach, {\it Phys. Rev.}
{\bf D61} (2000) 096002.
\medskip
\item{\sbl} S.-B. Liao, {\it Phys. Rev.} {\bf D53} (1996) 2020, and
{\it ibid.} {\bf D56} (1997) 5008.
\medskip
\item{\wave} S.-B. Liao, C.-Y. Lin and M. Strickland, in preparation. 
\medskip
\item{\dlitim} D. Litim, private communication.
\vfill 
\eject
\end